\documentstyle[aps]{revtex} 
\input amssym.def 
\input amssym.tex 
\def\m@th{\mathsurround=0pt}
\def\mymatrix#1{\null\,\vcenter{\normalbaselines\m@th
\ialign{\hfil$\scriptstyle ##$\hfil&&\quad\hfil $\scriptstyle
##$\hfil\crcr \mathstrut\crcr\noalign{\kern-\baselineskip}
#1\crcr\mathstrut\crcr\noalign{\kern-\baselineskip}}}\,}
\begin{document} 
\draft
\title{Supersymmetry for Systems with Unitary Disorder:
Circular Ensembles}
\author{Martin R. Zirnbauer${}^*$}
\address{Institute for Theoretical Physics, UCSB, Santa Barbara, U.S.A.}
\date{May 31, 1996}
\maketitle
\begin{abstract}
  A generalized Hubbard-Stratonovitch transformation relating an
  integral over random unitary $N \times N$ matrices to an integral
  over Efetov's unitary $\sigma$-model manifold, is introduced.  This
  transformation adapts the supersymmetry method to disordered and
  chaotic systems that are modeled not by a Hamiltonian but by their
  scattering matrix or time-evolution operator.  In contrast to the
  standard method, no saddle-point approximation is made, and no
  massive modes have to be eliminated.  This first paper on the
  subject applies the generalized Hubbard-Stratonovitch transformation
  to Dyson's Circular Unitary Ensemble.  It is shown how to use a
  supersymmetric variant of the Harish-Chandra-Itzykson-Zuber formula
  to compute, in the large-$N$ limit, the $n$-level correlation
  function for any $n$.  Nontrivial applications to random network
  models, quantum chaotic maps, and lattice gauge theory, are
  expected.
\end{abstract}
%\vfill\noindent
\bigskip
${}^*$ Permanent address: Institut f\"{u}r Theoretische Physik, 
Universit\"{a}t zu K\"{o}ln, Germany

\section{Introduction}

The long wave length physics of disordered metals is governed by the
so-called diffuson and cooperon modes describing the coherent
long-range propagation of particles and holes.  An efficient way to
handle the weak-disorder perturbation expansion involving these modes
is provided by the nonlinear $\sigma$-model representation pioneered
by Wegner\cite{wegner} and turned into a nonperturbative tool by
Efetov\cite{efetov}.  The great advantage of Efetov's supersymmetric
version of the $\sigma$ model is that it applies not only to diffusive
systems but also to systems in the ergodic and quantum limits and in
the localized regime.  In an exciting recent development, the
nonlinear $\sigma$ model was extended also to ballistic
systems\cite{mk} and to deterministic chaotic systems\cite{aasa}.

As it stands, Efetov's method is applicable to disordered systems that
are modeled by a Hamiltonian and where the disorder in the
Hamiltonian is weak and close to Gaussian.  It was not known until
very recently how to extend it to a closely related class of models
defined by a {\it unitary} operator rather than the Hamiltonian.  One
prominent example are random network models for, say, the random flux
problem or the percolation transition in integer quantum Hall
systems\cite{lc}.  These are formulated in terms of random scattering
matrices connecting incoming to outgoing channels at every node of a
network.  While there has been some progress in understanding the
Hamiltonian (or anisotropic) limit of these models, the original {\it
isotropic} models have so far defied analytical treatment.  A second
example of importance are periodically driven systems such as the
quantum kicked rotor\cite{izrailev} or the quantum kicked
top\cite{haake}.  Some of these systems have been found to possess
quasi-energy level statistics of a universal type given by Dyson's
circular random-matrix ensembles.  Others exhibit the quantum
phenomenon of dynamical localization, which corresponds to the
Anderson localization of disordered metallic wires, but a rigorous
derivation of the correspondence is as yet lacking.

In the present paper I am going to introduce a novel technique, akin
to the supersymmetry method of Efetov but different from it in its
details, which will allow analytical progress to be made with all of
the above models.  The heart of the formalism is a kind of generalized
Hubbard-Stratonovitch transformation that takes an integral over
random unitary $N\times N$ matrices (of the type that appears in the
Gaussian integral representation of a two-particle Green's function
for noninteracting electrons) and trades it directly for an integral
over Efetov's $\sigma$-model space of unitary symmetry.  The
transformation is exact for all $N$ including $N = 1$, no saddle-point
approximation is made, and no massive modes need to be eliminated.
The present paper is fairly technical as it aims to give a detailed
exposition of the method along with the necessary proofs.  The method
is illustrated at the simple example of Dyson's Circular Unitary
Ensemble (CUE).  Other, less trivial applications that have already
been worked out and are about to be published\cite{az_rotor,mrz_cc},
treat the quantum kicked rotor and the network model of Chalker and
Coddington.

The paper is organized as follows.  In Sec.~\ref{sec:HS} the
generalized Hubbard-Stratonovitch transformation is introduced in the
context of the calculation of the generating function for spectral
correlations of the CUE.  An integral over ${\rm U}(N)$ is replaced by
an integral over a pair of supermatrices $Z , \tilde Z$.  The
geometric meaning of these supermatrices is elucidated in
Sec.~\ref{sec:geometry}, where it is argued that they are coset
representatives for Efetov's coset space $G / H$ of unitary symmetry.
In the Appendix it is shown that the matrix elements of $Z$ can be
viewed as the parameters of a generalized coherent state obtained by
acting with Efetov's symmetry group $G$ (or, rather, its
complexification) on a certain lowest-weight module.  By embedding
this module into a Fock space of bosons and fermions we succeed in
giving an easy proof of the generalized Hubbard-Stratonovitch
transformation.  In Sec.~\ref{sec:largeN} the $n$-level correlation
function of the CUE is calculated for all $n$, by using a
supersymmetric version of the Harish-Chandra-Itzykson-Zuber formula.
Dyson's Circular Orthogonal Ensemble and the Circular Ensemble of type
$C$ are discussed in Secs.~\ref{sec:COE} and \ref{sec:type_C},
demonstrating the versatility of the new method.  A summary and
outlook is given in Sec.~\ref{sec:conclusion}.

The present work was inspired by unpublished results of Alexander
Altland, who had developed a similar (but less useful)
Hubbard-Stratonovitch scheme.  I am grateful to him for many
discussions of this subject.

\section{Circular Unitary Ensemble (type A)}
\label{sec:type_A}

Consider ${\rm U}(N)$, the unitary group in $N$ dimensions.  This
group, as any unimodular Lie group, comes with a natural integration
measure, the so-called Haar measure $dU$, which is invariant under
left and right translations $U \mapsto U_L U U_R$.  We normalize $dU$
by $\int_{{\rm U}(N)} dU = 1$.

The set of random unitary $N\times N$ matrices $U$ with probability
distribution given by the Haar measure $dU$, is called
Dyson's\cite{dyson} Circular Unitary Ensemble (CUE) in $N$ dimensions.
The CUE plays a central role in the random-matrix modeling of chaotic
and/or disordered systems with broken time-reversal symmetry.  A new
analytical method for treating the CUE is introduced below.  For
simplicity, the method will be illustrated at the example of the
generating function
        \begin{equation}
        F_{n_+ , n_-}( \{ \theta , \varphi \} ) = \int_{{\rm U}(N)} dU \ 
        \prod_{\alpha=1}^{n_+} {{\rm Det}(1-e^{i\varphi_{+\alpha}}U) 
        \over {\rm Det}(1 - e^{i\theta_{+\alpha}}U) }
        \prod_{\beta=1}^{n_-} {{\rm Det}(1-e^{-i\varphi_{-\beta}}
        U^\dagger) \over {\rm Det}(1 - e^{-i\theta_{-\beta}}U^\dagger) } ,
        \label{spec_det}
        \end{equation}
from which all information about the spectral correlations
of the CUE can be extracted.  $\varphi_{+\alpha}$,
$\varphi_{-\beta}$ and $\theta_{+\alpha}$, $\theta_{-\beta}$
are angular variables.  (The latter are given a small
imaginary part to prevent the determinants in the denominator
from becoming singular.)  The method does {\it not} exploit
the invariance of $F_{n_+ , n_-}$ under conjugation $U \mapsto
g U g^{-1}$, nor does it make any reference to the eigenbasis
of $U$.  For this reason it can be easily extended to the
calculation of more general types of correlation function,
such as those relating to wave functions and transport
coefficients.

\subsection{Generalized Hubbard-Stratonovitch Transformation}
\label{sec:HS}

The method we will use is a variant of Efetov's supersymmetry
technique\cite{efetov}, which has become a standard tool for 
treating Hamiltonian systems with Gaussian disorder.  Efetov's
method starts by the usual trick of expressing the ratio of 
determinants as a Gaussian superintegral:
        \begin{eqnarray}
        &&F_{n_+ , n_-}(\{ \theta , \varphi \} ) 
        \nonumber \\
        &=& \int_{{\rm U}(N)} dU \int D(\phi,\bar\phi;\chi,\bar\chi) 
        \ \exp \Big( - \bar\phi_{+\alpha}^k (\delta^{kl} 
        - e^{i\theta_{+\alpha}} U^{kl}) \phi_{+\alpha}^l 
        - \bar\chi_{+\alpha}^k (\delta^{kl} 
        - e^{i\varphi_{+\alpha}} U^{kl}) \chi_{+\alpha}^l 
        \nonumber \\
        &&\hspace{5cm} - \bar\phi_{-\beta}^k (\delta^{kl} - 
        e^{-i\theta_{-\beta}} \bar U^{lk}) \phi_{-\beta}^l 
        - \bar\chi_{-\beta}^k (\delta^{kl} - 
        e^{-i\varphi_{-\beta}} \bar U^{lk}) \chi_{-\beta}^l \Big) ,
        \nonumber
        \end{eqnarray}
where $\phi_{+\alpha}^k$, $\bar\phi_{+\alpha}^k$, $\phi_{-\beta}^l$,
and $\bar\phi_{-\beta}^l$ ($k,l = 1, ..., N$; $\alpha = 1, ..., n_+$;
$\beta = 1, ..., n_-$) are complex commuting (or bosonic) variables,
with the bar denoting complex conjugation, and $\chi,\bar\chi$ are
anticommuting (or fermionic) variables.  $U^{kl}$ are the matrix
elements of the unitary $N\times N$ matrix $U$.  The symbol
$D(\phi,\bar\phi; \chi,\bar\chi)$ denotes the ``flat'' Berezin
integration measure, i.e.  it is given by the product of the
differentials of the commuting variables times the product of the
partial derivatives with respect to the anticommuting variables.  The
summation convention is used throughout this paper.  The integral
converges for
        \[
        {\rm Im}\theta_{+\alpha} > 0 > {\rm Im}\theta_{-\beta} . 
        \]
To simplify the notation we introduce a composite index $a \equiv 
(\alpha,\sigma)$ where $\sigma$ takes values $\sigma = {\rm B}$ (Bosons) 
or $\sigma = {\rm F}$ (Fermions).  We then define a tensor $\psi$ with 
$(n_- + n_+)N =: nN$ components $\psi_{\pm a}^i$ by
        \[
        \psi_{\pm (\alpha,{\rm B})}^i = \phi_{\pm \alpha}^i, \quad
        \psi_{\pm (\alpha,{\rm F})}^i = \chi_{\pm \alpha}^i .
        \]
$\bar\psi$ is defined similarly.  Setting $D(\phi,\bar\phi;\chi,\bar\chi)
\equiv D(\psi,\bar\psi)$ and combining the phases $\{ \theta , \varphi
\}$ into a single symbol $\{ \omega \}$ by
        \[
        \omega_{\pm (\alpha,{\rm B})} = \theta_{\pm\alpha}, \quad
        \omega_{\pm (\alpha,{\rm F})} = \varphi_{\pm\alpha} ,
        \]
we may express the generating function in the abbreviated form
        \begin{eqnarray}
        F_{n_+ , n_-}(\{ \omega \} ) = 
        \int D(\psi,\bar\psi) \int_{{\rm U}(N)} dU
        \ \exp \Big( &&- \bar\psi_{+a}^k (\delta^{kl} - e^{i\omega_{+a}}
        U^{kl}) \psi_{+a}^l 
        \nonumber \\
        &&- \bar\psi_{-b}^k (\delta^{kl} - 
        e^{-i\omega_{-b}} \bar U^{lk}) \psi_{-b}^l \Big) .
        \label{gen_func}
        \end{eqnarray}

In the case of Hamiltonian systems, the next steps of Efetov's method
are (i) to average over the Gaussian disorder, thereby producing a 
term quartic in the $\psi$ fields, (ii) to decouple this term by the
introduction of a Hubbard-Stratonovich supermatrix field $Q$, and
(iii) to make a saddle-point approximation eliminating the so-called
massive modes.  The final outcome of this convoluted and
mathematically very delicate procedure is an integral over fields that
live on a nonlinear supermanifold, called Efetov's $\sigma$-model
space with unitary symmetry.  It is difficult to see how this
procedure could be adapted to the present case as the Haar integral
over ${\rm U}(N)$ is very far from being Gaussian.  Fortunately, there 
is another way to proceed.

The key innovation made here is to trade the integral over ${\rm
U}(N)$ {\it directly} for an integral over Efetov's $\sigma$-model
space with unitary symmetry.  This is achieved by the following
remarkable identity:
        \begin{eqnarray}
        &&\int_{{\rm U}(N)} dU \ \exp \left( 
        {\bar\psi}_{+a}^i U^{ij} \psi_{+a}^j +  
        {\bar\psi}_{-b}^j \bar U^{ij} \psi_{-b}^i \right)
        \nonumber \\
        = &&\int D\mu_N(Z,\tilde Z) \ \exp \left( \bar\psi_{+a}^i 
        Z_{ab}^{\vphantom{i}} \psi_{-b}^i + \bar\psi_{-b}^j 
        {\tilde Z}_{ba}^{\vphantom{i}} \psi_{+a}^j \right) ,
        \label{HS}
        \end{eqnarray}
which can be proved by letting the Lie supergroup ${\rm Gl}(nN|nN)$
act on a Fock space of bosons and fermions and using some basic
concepts of the theory of generalized coherent states.  The proof is
given in detail in the Appendix.  Here we confine ourselves to spelling 
out the meaning of the right-hand side.  The new integration variables
$Z = \{ Z_{ab}\}$ and $\tilde Z = \{ \tilde Z_{ab}\}$ are supermatrices
whose Boson-Fermion block decomposition is written
        \[
        Z = \pmatrix{ Z_{\rm BB} &Z_{\rm BF}\cr Z_{\rm FB} 
        &Z_{\rm FF}\cr}, \quad
        \tilde Z = \pmatrix{ \tilde Z_{\rm BB} &\tilde Z_{\rm BF}\cr 
        \tilde Z_{\rm FB} &\tilde Z_{\rm FF}\cr}.
        \]
Each block of $Z$ ($\tilde Z$) is a rectangular matrix of dimension
$n_+ \times n_-$ $(n_- \times n_+)$.  The integration measure is
        \[
        D\mu_N(Z,\tilde Z) = D(Z,\tilde Z) \ {\rm SDet}(1-\tilde Z Z)^N ,
        \]
where $D(Z,\tilde Z)$ denotes the flat Berezin measure on the space of
supermatrices $Z,\tilde Z$,\footnote{To be precise, I should say that
this is true only {\it locally}.  Berezin measures typically suffer
from a global anomaly\cite{rothstein2,mrz_suprev}.  What determines
$D(Z,\tilde Z)$ and its anomaly uniquely (up to multiplication by a
constant) is the requirement of invariance under translations on
Efetov's $\sigma$-model space $G / H$, see Sec.~\ref{sec:geometry} and
the Appendix.}  and ${\rm SDet}$ is the superdeterminant.  The domain 
of integration is defined by the conditions
        \[
        \tilde Z_{\rm BB}^{\vphantom{\dagger}} = Z_{\rm BB}^\dagger , \quad
        \tilde Z_{\rm FF}^{\vphantom{\dagger}} = - Z_{\rm FF}^\dagger ,
        \]
and the requirement that all eigenvalues of the positive hermitian
$n_- \times n_-$ matrix $\tilde Z_{\rm BB} Z_{\rm BB}$ be less than
unity.  The integration measure is normalized by $\int D\mu_N(Z,\tilde
Z) = 1$.  In the next subsection we will argue that $Z$ and  $\tilde Z$ 
should be interpreted as parameterizing Efetov's unitary $\sigma$-model 
space.

Let me mention in passing that integrals of a type similar to the
left-hand side of (\ref{HS}) appear in lattice gauge theory, where
$\psi$ and $U$ represent the quark and gluon fields, respectively.
Motivated by this similarity, one might refer to the upper index of
$\psi$ as ``color'' and the lower one as ``flavor''.  We could then
say that (\ref{HS}) integrates out the gluon fields, which carry
color, and replaces them by an integral over color-singlet meson
fields carrying flavor.  Note that the color degrees of freedom of the
tensor $\psi$ are {\it uncoupled} on the right-hand side.

By using the generalized Hubbard-Stratonovitch transformation (\ref{HS}) 
with $\bar\psi_{+a}^k \to \bar\psi_{+a}^k e^{i\omega_{+a}}$ and 
$\bar\psi_{-b}^k \to \bar\psi_{-b}^k e^{-i\omega_{-b}}$, we can process 
the expression (\ref{gen_func}) as follows:
        \begin{eqnarray}
        F_{n_+ , n_-} (\{ \omega \}) &=& 
        \int D\mu_N(Z,\tilde Z) \int D(\psi,\bar\psi)
        \ \exp \Big( - \bar\psi_{+a}^k \psi_{+a}^k - \bar\psi_{-b}^k
        \psi_{-b}^k
        \nonumber \\
        &&\hspace{6cm} + \bar\psi_{+a}^k e^{i\omega_{+a}} 
        Z_{ab}^{\vphantom{i}} \psi_{-b}^k + \bar\psi_{-b}^k 
        e^{-i\omega_{-b}} \tilde Z_{ba}^{\vphantom{i}}\psi_{+a}^k \Big)
        \nonumber \\
        &=& \int D\mu_N(Z,\tilde Z) \ {\rm SDet}^{-N}
        \pmatrix{1 &-e^{i\omega_+}Z\cr -e^{-i\omega_-}\tilde Z &1\cr}
        \nonumber \\
        &=& \int D\mu_N(Z,\tilde Z) \ {\rm SDet}^{-N}
        \left( 1 - \tilde Z e^{i\omega_+} Z e^{-i\omega_-} \right)
        \nonumber \\
        &=& \int D(Z,\tilde Z) \ {\rm SDet}^{-N}
        \left( 1 - (1-\tilde Z Z)^{-1} \tilde Z ( e^{i\omega_+} Z 
        e^{-i\omega_-} - Z ) \right) .
        \label{fin_result}
        \end{eqnarray}
In the first step we did the Gaussian superintegral over $\psi,
\bar\psi$, producing a superdeterminant.  As the different 
species labeled by $i = 1,...,N$ are uncoupled, this superdeterminant
separates into a product of $N$ factors.  The expression
$e^{i\omega_+}Z$ stands for the supermatrix with matrix elements
$e^{i\omega_{+a}}Z_{ab}$.  The second step in (\ref{fin_result}) is
immediate from the elementary identity
        \[
        {\rm SDet} \pmatrix{ A &B\cr C &D\cr} = 
        {\rm SDet}(A) \ {\rm SDet}(D-CA^{-1}B) .
        \]
The third step follows from the formula $D\mu_N(Z,\tilde Z) = D(Z,
\tilde Z) \ {\rm SDet}^N(1-\tilde Z Z)$ along with the multiplicative
property of the superdeterminant: ${\rm SDet}(A) \ {\rm SDet}(B) =
{\rm SDet}(AB)$.

Let me emphasize that the result (\ref{fin_result}) is {\it exact for
all $N$}.  In contrast with Efetov's method, {\it no saddle-point
approximation was used, and no massive modes had to be eliminated}.
Equation (\ref{fin_result}) reduces the integral (\ref{spec_det}) over
the $N^2$ real freedoms of ${\rm U}(N)$ to an integral over the $4n_+
n_-$ complex freedoms of the supermatrices $Z, \tilde Z$.  Clearly,
such a scheme is efficient when $N$ is large and $n_+ n_-$ is small.

\subsection{Geometric meaning of $Z, \tilde Z$}
\label{sec:geometry}

We start with a concise mathematical description of Efetov's
$\sigma$-model space of unitary symmetry.  Let $G= {\rm Gl}(n|n)$, the
complex Lie supergroup of regular supermatrices of dimension
$(n+n)\times (n+n)$, and put $n = n_A + n_R$.  (Later we shall make
the identifications $n_A = n_+$ and $n_R = n_-$.)  Using
tensor-product notation, we take $\Sigma_z \in {\rm Gl}(n|n)$ to be
the diagonal matrix $\Sigma_z = 1_{1|1} \otimes {\rm diag} (1_{n_A},
-1_{n_R})$ where $1_{n_A}$ is the $n_A$-dimensional unit matrix, and
$1_{1|1}$ is the $(1+1)$-dimensonal unit matrix in Boson-Fermion
space.  Matrices $h \in {\rm Gl}(n|n)$ that commute with $\Sigma_z$
are of the form
        \[
        h = \pmatrix{h_+ &0\cr 0 &h_-\cr}, \qquad {\rm if} \quad
        \Sigma_z = \pmatrix{1_{n_A|n_A} &0\cr 0 &-1_{n_R|n_R}\cr}
        \]
is the matrix presentation of $\Sigma_z$.  Thus the condition $h
\Sigma_z = \Sigma_z h$ (or, equivalently, $h = \Sigma_z h \Sigma_z$)
fixes a subgroup $H = {\rm Gl}(n_A|n_A) \times {\rm Gl}(n_R|n_R)$ of
$G$.  The coset space $G / H $ is a complex-analytic supermanifold in
the sense of Berezin-Kostant-Leites\cite{bl,kostant}, and has complex
dimension $d_{\rm Boson} = d_{\rm Fermion} = 2 n_A n_R$.  The
canonical projection $G \to G/H$ endows the coset space $G/H$ with a
natural $G$-invariant geometry.  If $G/H$ is modeled by supermatrices
$Q = g \Sigma_z g^{-1}$, this geometry is given by the rank-two tensor
field ${\rm STr}({\rm d}Q)^2$, where ${\rm STr}$ denotes the
supertrace.  The $G$-invariant Berezin integration measure on $G / H$
is denoted by $Dg_H$, or $DQ$.  The objects one wants to
integrate\cite{mrz_suprev} are the holomorphic functions on $G/H$,
i.e. functions with a holomorphic dependence on a set of complex local
coordinates of the complex-analytic supermanifold $G / H$.  Such
functions are written $f(gH)$ or $F(Q)$.  The domain of integration is
taken to be a Riemannian submanifold $M_{\rm B} \times M_{\rm F}$ of
the support of $G / H$, where
        \begin{eqnarray}
        M_{\rm B} &=& {\rm U}(n_A,n_R) / {\rm U}(n_A) \times {\rm U}(n_R)
        \qquad ({\rm BB~sector}), \nonumber \\
        M_{\rm F} &=& {\rm U}(n_A+n_R) / {\rm U}(n_A) \times {\rm U}(n_R)
        \qquad ({\rm FF~sector}). \nonumber
        \end{eqnarray}
In this way one gets a $G$-invariant Berezin integral
        \begin{equation}
        \int_{M_{\rm B} \times M_{\rm F}} Dg_H \ f(gH)
        = \int_{M_{\rm B} \times M_{\rm F}} Dg_H \ f(g_0 gH)
        \qquad (g_0 \in G),
        \label{inv_int}
        \end{equation}
which is called the integral over Efetov's $\sigma$-model space with
unitary symmetry.  Translation $gH \mapsto g_0 gH$ by an
element $g_0 \in G$ does not leave the integration domain $M_{\rm B}
\times M_{\rm F}$ invariant, in general.  Nevertheless, if $f$ is
holomorphic, as is assumed, (\ref{inv_int}) does hold because Cauchy's
theorem applies and allows to undo the deformation of the
integration domain.  In the terminology of\cite{mrz_suprev} the
pair $(G/H,M_{\rm B}\times M_{\rm F})$ is called a Riemannian
symmetric superspace of type $A{\rm III}|A{\rm III}$.  (Nine more
types exist.)

To make contact with Sec.~\ref{sec:HS} we introduce a suitable
parameterization of $G/H$.  Using the above presentation where
$\Sigma_z = {\rm diag}(1_{n_A|n_A} , - 1_{n_R|n_R})$ we decompose $g
\in G$ as
        \[      
        g = \pmatrix{A &B \cr C &D \cr} .
        \]
Since $h = {\rm diag}(h_+ , h_-)$, the invariants for the right action
of $h \in H$ on $G$ are $Z := BD^{-1}$ and $\tilde Z := CA^{-1}$.  We
may take the matrix elements of $Z, \tilde Z$ for a set of complex
local coordinates of the coset space $G / H$.  The expression for
$Q = g \Sigma_z g^{-1}$ in these coordinates is 
        \begin{equation}
        Q = \pmatrix{1 &Z\cr \tilde Z &1\cr} \pmatrix{1 &0\cr 0 &-1\cr}
        \pmatrix{1 &Z\cr \tilde Z &1\cr}^{-1} . 
        \label{Q_by_Z}
        \end{equation}
From the expression for the invariant rank-two tensor field, 
        \[
        {\rm STr}({\rm d}Q)^2 = {\rm const} \times
        {\rm STr}(1-\tilde Z Z)^{-1} {\rm d}\tilde Z (1-Z \tilde Z)^{-1}
        {\rm d}Z ,
        \]
one easily finds the $G$-invariant Berezin measure $DQ$ in these coordinates
to be the (locally) flat one, $D(Z,\tilde Z)$\cite{haw_mrz}.  The flatness 
results from cancellations due to supersymmetry. 

To make the restriction to the Riemannian submanifold $M_{\rm B}
\times M_{\rm F}$, note that for a unitary matrix $g = \left(
\mymatrix{a &b\cr c &d\cr} \right) \in {\rm U}(n_A + n_R)$, we have
$b^\dagger = - (d - ca^{-1}b)^{-1}ca^{-1}$ and $d^\dagger = (d - c
a^{-1} b)^{-1}$.  This implies
        \[
        Z_{\rm FF}^\dagger = (bd^{-1})^\dagger = {d^{-1}}^\dagger 
        b^\dagger = - c a^{-1} = - \tilde Z_{\rm FF} .
        \]
For the noncompact case $g \in {\rm U}(n_A,n_R)$, the expression for
$b^\dagger$ has its sign reversed, so
        \[
        Z_{\rm BB}^\dagger = {d^{-1}}^\dagger b^\dagger = 
        + c a^{-1} = + \tilde Z_{\rm BB} .
        \]
Moreover, the pseudo-unitarity of $g \in {\rm U}(n_A,n_R)$ implies
that $a^\dagger a - c^\dagger c = 1$, from which we infer
        \[
        \tilde Z_{\rm BB} Z_{\rm BB} = c a^{-1} {a^{-1}}^\dagger c^\dagger
        = c (1 + c^\dagger c)^{-1} c^\dagger ,
        \]
so that all eigenvalues of the positive hermitian $n_R \times n_R$
matrix $\tilde Z_{\rm BB} Z_{\rm BB}$ must be less than unity.

All these facts suggest to put $n_A \equiv n_+$, $n_R \equiv n_-$ and
identify the supermatrices $Z, \tilde Z$ defined here with those
figuring in the generalized Hubbard-Stratonovitch transformation
(\ref{HS}).  This identification is made rigorous in the Appendix,
where a Lie-algebraic proof of (\ref{HS}) is given.  The proof relies
in an essential way on the existence of the invariant integral
(\ref{inv_int}), and on the identification of $D(Z,\tilde Z)$ with the
$G$-invariant Berezin measure $Dg_H$ on $G / H$.

For further orientation, consider the simplest example $n_+ = n_- =
1$, where $M_{\rm B} = {\rm U}(1,1)/{\rm U}(1)\times{\rm U}(1) \simeq
{\rm H}^2$ (two-hyperboloid) and $M_{\rm F} = {\rm U}(2)/{\rm U}(1)
\times {\rm U}(1) \simeq {\rm S}^2$ (two-sphere).  In this case the
BB and FF blocks of $Z$ are just numbers.  The expression for $Z_{\rm
  FF}$ in terms of the usual polar $(\theta)$ and azimuthal
$(\varphi)$ angles on ${\rm S}^2$ turns out\cite{mrz_iqhe} to be
$Z_{\rm FF} = \tan( \theta/2) \exp i\varphi$.  From this it is easily
seen that $Z_{\rm FF}$ can be interpreted as being a complex
stereographic coordinate for ${\rm S}^2$.  Its BB analog can be
parameterized in terms of the hyperbolic polar angle $\theta_{\rm B}$
on ${\rm H}^2$ by $Z_{\rm BB} = \tanh (\theta_{\rm B}/2) \exp
i\varphi_{\rm B}$.

To conclude this subsection, let us recast the result (\ref{fin_result})
in the coordinate-free language developed here.  From (\ref{Q_by_Z}) or
        \[
        Q = g \Sigma_z g^{-1} = 
        \pmatrix{1 &Z\cr \tilde Z &1\cr}
        \pmatrix{(1-Z\tilde Z)^{-1} &-Z(1-\tilde Z Z)^{-1}\cr
        {\tilde Z}(1-Z\tilde Z)^{-1} &-(1-\tilde Z Z)^{-1}\cr}
        =: \pmatrix{Q_{++} &Q_{+-}\cr Q_{-+} &Q_{--}\cr} ,
        \]
we read off the relations
        \begin{eqnarray}
        &&\tilde Z(1-Z\tilde Z)^{-1} = Q_{-+} / 2 , \quad
        (1-Z\tilde Z)^{-1} = (1+Q_{++}) / 2 ,
        \nonumber \\
        &&Z = - (1+Q_{++})^{-1} Q_{+-} , \quad
        (1-\tilde Z Z)^{-1} = (1-Q_{--}) / 2 .
        \nonumber
        \end{eqnarray}
Inserting these into (\ref{fin_result}) we can express the
generating function as a $Q$-integral:
        \begin{equation}
        F_{n_+ , n_-}(\{ \omega \}) = \int DQ \ {\rm SDet}^{-N} \left(
        1 - Q_{--} + Q_{-+} e^{i\omega_+} (1+Q_{++})^{-1} Q_{+-}
        e^{-i\omega_-} \right) .
        \label{Q_integral}
        \end{equation}
Let me repeat that this expression is exact for all $N$.

The following remark may be helpful.  It is well known\cite{mehta}
that the density of states of the {\it Gaussian} Unitary Ensemble
tends to a semicircle in the large-$N$ limit.  It is also known\cite{vwz}
that the so-called ``massive modes'', which show up in Efetov's method
and are eliminated by a saddle-point approximation, are needed to
reproduce the {\it nonuniformity} of the semicircular density of
states.  In contradistinction, the eigenvalues of a unitary matrix $U
\in {\rm CUE}$ are uniformly distributed over the unit circle, on
average.  Qualitatively speaking, it is this {\it uniformity} of the
CUE spectrum that allows a $Q$-integral representation to be derived
without introducing the massive modes and without making any
saddle-point approximation.

\subsection{Large-N Limit}
\label{sec:largeN}

For further calculations the formula (\ref{fin_result}) turns out to
be most convenient.  In the large-$N$ limit the integrand in
(\ref{fin_result}) substantially differs from zero only when the
difference of matrices $\omega_+ - \omega_-$ is of order $1/N$.  This
fact allows us to expand:
        \[
        e^{i\omega_+} Z e^{-i\omega_-} - Z = i \left(
        \omega_+ Z - Z \omega_- \right) + {\cal O}(1/N^2) ,
        \]
which leads to the approximate equality
        \[
        F_{n_+ , n_-} ( \{ \omega \} ) \simeq 
        \int D(Z,\tilde Z) \ \exp iN {\rm STr} 
        \left( \omega_+ Z\tilde Z (1-Z \tilde Z)^{-1} - \omega_- 
        \tilde Z Z(1-\tilde Z Z)^{-1} \right) .
        \nonumber
        \]
The $N$ eigenvalues of the unitary matrix $U$ are distributed over the
unit circle with circumference $2\pi$, which results in the mean density 
of eigenphases being $\nu = N / 2\pi$.  By setting $\omega := {\rm diag} (
\omega_+ , \omega_- )$ we can write the integrand in the form 
$\exp iN {\rm STr} \omega (Q-\Sigma_z) / 2$.  Thus, the large-$N$ limit 
of the generating function $F_{n_+ , n_-} (\omega \equiv \{ \omega \})$ 
evaluated on a scale set by the mean spacing, is
        \[
        \lim_{N\to\infty} F_{n_+ , n_-} (2\pi \omega / N) = \int dQ
        \exp i \pi {\rm STr} \omega ( Q - \Sigma_z ) .
        \]
Equivalently, we may write $Q = g \Sigma_z g^{-1}$, and get
        \begin{equation}
        \lim_{N\to\infty} F_{n_+ , n_-} (2\pi \omega / N) = 
        \int_{M_{\rm B}\times M_{\rm F}} Dg_H 
        \exp i\pi {\rm STr} \omega ( g \Sigma_z g^{-1} - \Sigma_z ) .
        \label{fin_res_g}
        \end{equation}
As is well known\cite{efetov,vwz,mrz_suprev}, the same expression is
obtained for the Gaussian Unitary Ensemble, which confirms the expected
equivalence\cite{dyson} of the Circular and Gaussian ensembles in the
large-$N$ limit.

We are now going to calculate from the large-$N$ result
(\ref{fin_res_g}) the so-called $n$-level correlation function $R_n$
for any $n$. (Efetov\cite{efetov} calculated only the two-level
correlation function.) To that end, we will use a supersymmetric
variant of a result known in the physics literature as the
Itzykson-Zuber formula\cite{iz}, which is actually a special case of a
more general result proved 23 years earlier by the mathematician
Harish-Chandra.

Let $K$ be any connected compact semisimple Lie group, which we
assume to be realized by $n \times n$ matrices $k$, and let ${\cal
T}$ be a maximal commuting (or Cartan) subalgebra of ${\rm Lie}(K)$.
Without loss we take ${\cal T}$ to be the diagonal matrices in ${\rm
Lie} (K)$.  If $A, B$ are two elements of ${\cal T}$, Harish-Chandra
showed that (see Theorem 2 of~\cite{harish-chandra})
        \begin{equation}
        \int_{K} dk \exp {\rm Tr} A k B k^{-1} = 
        {{\rm const}\over p(A) p(B)} \sum_{\hat s \in {\rm W}[K]}
        (-1)^{|\hat s|} \exp {\rm Tr} A {\hat s} B ,
        \label{harish-chandra}
        \end{equation}
where $p(A) = \prod_{\alpha > 0} \alpha(A)$ is the product over a set
of positive roots $\alpha$ of the pair $[ {\cal T}, {\rm Lie}(K) ]$,
and ${\rm const}$ is a normalization constant.  The sum runs over the
Weyl group ${\rm W}[K]$, which is the discrete group of inequivalent
transformations ${\cal T} \to {\cal T}$ by $A \mapsto {\hat s} A := s
A s^{-1}$ with $s \in K$.  The Weyl group can be considered as a
subgroup of the symmetric group ${\rm S}_n$ acting on the elements of 
the diagonal matrices $A \in {\cal T}$.  The symbol $|\hat s| = 0, 1$
denotes the parity of $\hat s$.  For the special case $K = {\rm
  SU}(n)$ (or ${\rm U}(n)$, it makes no real difference) one has $p(A) =
\prod_{i < j} (A_i - A_j)$, where $A_i$ are the elements of the
diagonal matrix $A$, and the Weyl group coincides with ${\rm S}_n$,
i.e. $(\hat s A)_i = (s A s^{-1})_i =: A_{{\hat s}(i)}$ where $\hat s
\in {\rm S}_n$ is a permutation and $|\hat s|$ is the parity of that
permutation.  The general result (\ref{harish-chandra}) thus takes the
more familiar form\cite{iz}
        \[
        \int_{{\rm U}(n)} dU \exp {\rm Tr} A U B U^{-1} = 
        {\rm const} \times
        \prod_{i<j} (A_i - A_j)^{-1} (B_i - B_j)^{-1} \times
        {\rm Det} \left( e^{A_i B_j} \right)_{i,j=1,...,n} .
        \]

The right-hand side of (\ref{harish-chandra}) can be viewed as the
stationary-phase approximation to the integral on the left-hand side,
with the points of stationarity being enumerated by the elements $\hat
s \in {\rm W} [ K ]$.  Thus, Harish-Chandra's formula states that the
stationary-phase approximation is {\it exact} in this case.
M.~Stone\cite{mstone} has traced this remarkable coincidence to a
hidden symmetry of the integral.

I conjecture that the formula (\ref{harish-chandra}) extends to any 
compact classical Lie {\it super}group $K$:
        \begin{equation}
        \int_{K} Dk \ \exp {\rm STr} A k B k^{-1}
        = {{\rm const}\over p(A) p(B)} \sum_{\hat s \in {\rm W}[K_0]}
        (-1)^{|\hat s|} \exp {\rm STr} A \hat s B ,
        \label{super_hc}
        \end{equation}
where ${\rm W}[K_0]$ now is the Weyl group of the ordinary group $K_0$
supporting $K$, and $p(A)$ turns into the product of positive bosonic
roots divided by the product of positive fermionic roots.  The
strategy of the proof ought to be a supersymmetric extension of that of 
Harish-Chandra and use the theory of invariant differential
operators\cite{harish-chandra,helgason,fuchs}.  For the special case
of the unitary Lie supergroup $K = {\rm U}(n|n)$, which is the one we
will need below, two proofs using alternative strategies can be found
in Refs.~\cite{guhr2} and\cite{amu}.  In that case, if 
$A = {\rm diag}(A_{1,{\rm B}} , ... , A_{n,{\rm B}} ; A_{1,{\rm F}} ,
... , A_{n,{\rm F}})$, one has
        \begin{equation}
        p(A) = \prod_{i < j}(A_{i,{\rm B}}-A_{j,{\rm B}}) (A_{i,{\rm
        F}}-A_{j,{\rm F}}) / \prod_{i,j}(A_{i,{\rm B}}-A_{j,{\rm F}}) ,
        \label{prod_roots}
        \end{equation}
and ${\rm W}[K_0] = {\rm W}[{\rm U}(n)\times{\rm U}(n)] \simeq 
{\rm S}_n \times {\rm S}_n$, so
        \[
        \sum_{\hat s \in {\rm W}[K_0]} (-1)^{|\hat s|} \exp {\rm STr} A 
        \hat s B = {\rm Det}\left( e^{A_{i,{\rm B}} B_{j,{\rm B}}}
        \right)_{i,j=1,...,n} \times {\rm Det}\left( e^{-A_{i,{\rm F}} 
        B_{j,{\rm F}}} \right)_{i,j=1,...,n} .
        \]
In a reduced form tailored to physical applications, the 
Harish-Chandra-Itzykson-Zuber integral for ${\rm U}(n|n)$ 
made its first appearance in\cite{guhr1}.

Now recall that the $n$-level correlation function $R_n(\theta_1,
\theta_2,...,\theta_n)$ is defined\cite{mehta} as the probability
density to find, given a level (i.e. an eigenphase) at $\theta_1$, $n
- 1$ levels at positions $\theta_2, ..., \theta_n$, irrespective of
the positions of all other levels.  By making use of the identities
        \begin{eqnarray}
        &&{\partial\over \partial\varphi}\Big|_{\varphi = \theta} { 
        {\rm Det}(1 - e^{i\varphi} U) \over {\rm Det} (1 - e^{i\theta} U)}
        = i {\rm Tr} \left( 1 - e^{-i\theta} U^\dagger \right)^{-1} ,
        \nonumber \\
        &&{ e^{-\varepsilon} \varepsilon/\pi \over (1 - e^{-\varepsilon - 
        i\theta}) (1 - e^{-\varepsilon + i\theta}) } = { \varepsilon / \pi
        \over 2(1-\cos\theta) + 4 \sinh^2(\varepsilon/2)} \ {\buildrel 
        {\varepsilon\to 0} \over \longrightarrow} \ \delta(\theta) ,
        \nonumber
        \end{eqnarray}
we can extract $R_n$ from the generating function $F_{n,n}$, 
Eq.~(\ref{spec_det}), by the following formula:
        \begin{equation}
        R_n(\theta_1 , ... , \theta_n) = \lim_{\varepsilon\to 0}
        \left(\varepsilon \over \pi\right)^n \prod_{\alpha = 1}^n 
        {\partial^2\over\partial\varphi_{+\alpha}\partial\varphi_{-\alpha} 
        } \Big|_{\varphi_{\pm\alpha} = \theta_{\alpha} \pm i\varepsilon}
        F_{n,n}( \{ \theta_{\alpha}+i\varepsilon , \theta_{\alpha}-
        i\varepsilon , \varphi_{+\alpha},\varphi_{-\alpha} \} ) ,
        \label{extract}
        \end{equation}
which serves as the starting point of our supersymmetric calculation
of $R_n$. 

The first step toward doing the integral (\ref{fin_res_g}) is to
introduce Efetov's polar coordinates\cite{efetov} on the coset space
$G / H$:
        \begin{eqnarray}
        &&g H = h a H , \quad {\rm where} \quad 
        h = \pmatrix{ h_+ &0\cr 0 &h_-\cr} \in H ,
        \nonumber \\
        &&{\rm and} \quad
        a = \exp \pmatrix{0 &z\cr z &0\cr}, \quad
        z = {\rm diag}(x_1,...,x_n,iy_1,...,iy_n) .
        \nonumber
        \end{eqnarray}
The variables $x_i, y_j$ contained in $a$ are the ``radial''
coordinates, and the supermatrix $h$ is parameterized by a set of
``angular'' coordinates.  The choice of integration domain $M_{\rm B}
\times M_{\rm F}$ for $G / H$ dictates that we must integrate over
{\it real} $x_i , y_j$ and superunitary $h \in {\rm U}(n|n) \times
{\rm U}(n|n) \subset H$.  The polar-coordinate form of the invariant
integral (\ref{inv_int}) is
        \begin{equation}
        \int Dg_H f(gH) = \int \left( \int Dh \ f(h a H)
        \right) J(a) da + ... ,
        \label{polar_int}
        \end{equation}
where $Dh$ is the Haar-Berezin measure of $H$, and $da$ denotes the
Euclidean measure on the abelian group of radial elements $a$.  The
Jacobian of the transformation to polar coordinates is
given\cite{hz,fb} by $J(a)da = $
        \begin{eqnarray}
        &&{\prod_{i<j} \sinh^2(x_i - x_j) \sinh^2(x_i + x_j)
        \sin^2(y_i - y_j) \sin^2(y_i + y_j) \over
        \prod_{i,j} \sinh^2(x_i + i y_j) \sinh^2(x_i - i y_j)}
        \prod_{k=1}^n \sinh(2x_k) \sin(2y_k) {\rm d}x_k {\rm d}y_k 
        \nonumber \\
        &=& {\rm const} \times 
        {\prod_{i<j} (\cosh 2x_i - \cosh 2x_j)^2 (\cos 2y_i - \cos 2y_j)^2
        \over \prod_{i,j} (\cosh 2x_i - \cos 2y_j)^2}
        \prod_{k=1}^n {\rm d}(\cosh 2x_k) {\rm d}(\cos 2y_k) .
        \label{jacobian}
        \end{eqnarray}
For the second equality sign the trigonometric identity 
        \[
        2 \sinh(\alpha + \beta) \sinh(\alpha - \beta) = 
        \cosh 2\alpha - \cosh 2\beta
        \]
was used.  The integration domain is taken to be $0 \le x_k < \infty$,
$0 \le y_k \le \pi/2$.  The dots in (\ref{polar_int}) indicate
correction terms that arise from the anomaly of the invariant Berezin
measure in polar coordinates.  The form of these corrections was
investigated and completely determined in\cite{bundschuh}.  Their main
characteristic is that they are supported on the {\it boundary} of the
integration domain (which is why they are alternatively known as
``boundary terms'').  More precisely, they are obtained by setting
$x_i = y_j = 0$ for one or several pairs $(x_i,y_j)$.  It will turn out 
that these boundary terms do not contribute to $R_n$, Eq.~(\ref{extract}),
and therefore do not need to be specified explicitly for our purposes.

The next step is to express ${\rm STr}\omega g \Sigma_z g^{-1}$ in the
exponential of the integrand of (\ref{fin_res_g}) in terms of
polar coordinates:
        \begin{eqnarray}
        {\rm STr} \omega g\Sigma_z g^{-1} = {\rm STr} \Sigma_z \omega 
        h a^2 h^{-1} &=& {\rm STr} \Sigma_z \omega h \cosh(2\ln a) h^{-1}
        \nonumber \\
        &=& {\rm STr} \omega_+^{\vphantom{-1}} h_+^{\vphantom{-1}}
        A h_+^{-1} - {\rm STr} \omega_-^{\vphantom{-1}} 
        h_-^{\vphantom{-1}} A h_-^{-1} ,
        \nonumber
        \end{eqnarray}
where $A = \cosh 2z$, $\omega = {\rm diag}(\omega_+ , \omega_-)$, and
$\omega_\pm = {\rm diag}(\theta_1 \pm i\varepsilon , ... , \theta_n
\pm i\varepsilon ; \varphi_{\pm 1} , ... , \varphi_{\pm n})$.  
By this and $Dh = Dh_+ Dh_-$ the angular integral 
over ${\rm U}(n|n) \times {\rm U}(n|n)$ factors into a product of two
integrals over ${\rm U}(n|n)$.  By formula (\ref{super_hc}) 
for ${\rm U}(n|n)$, the integral over $h_+$ gives
        \[
        \int_{{\rm U}(n|n)} Dh_+ \exp i {\rm STr} \omega_+^{\vphantom{-1}}
        h_+^{\vphantom{-1}} A h_+^{-1} = {{\rm const}\over p(\omega_+) p(A)} 
        \sum_{\hat s} (-1)^{|{\hat s}|} \exp i {\rm STr} \omega_+ {\hat s} A .
        \]
A similar expression results from doing the integral over $h_-$.
Multiplying both integrals we get a factor $p(A)^{-2}$.
By (\ref{prod_roots}) and $A = {\rm diag}(\cosh 2x_1 , ... ,
\cosh 2x_n ; \cos 2y_1 , ... , \cos 2y_n )$ this factor exactly
cancels the factor multiplying $dA := \prod_k {\rm d}(\cosh 2x_k) 
{\rm d}(\cos 2y_k)$ in (\ref{jacobian}).  Hence we obtain
        \begin{eqnarray}
        &&\int Dg_H \exp i {\rm STr} \omega g \Sigma_z g^{-1} 
        \nonumber \\
        &=& {{\rm const}\over p(\omega_+) p(\omega_-)}\sum_{{\hat s}_+,
        {\hat s}_-} (-1)^{|{\hat s}_+| + |{\hat s}_-|} \int dA \exp i 
        {\rm STr} A \left( {\hat s}_+^{\vphantom{-1}} \omega_+^{\vphantom{-1}}
        - {\hat s}_-^{\vphantom{-1}} \omega_-^{\vphantom{-1}} \right) 
        \nonumber \\
        &&\hspace{9cm} + \ {\rm boundary} \ {\rm terms} . \nonumber
        \end{eqnarray}
The inverse of the product of positive roots $p(\omega_+)^{-1}$
vanishes linearly with each difference $\omega_{+\alpha,{\rm B}} -
\omega_{+\alpha,{\rm F}} = \theta_{\alpha} +i\varepsilon - \varphi_{
+\alpha}$.  An analogous statement holds for $p(\omega_-)^{-1}$.
Therefore, on taking the $2n$ first derivatives at $\varphi_{\pm\alpha} =
\theta_{\alpha} \pm i\varepsilon$ [see (\ref{extract})], we
simply obtain a constant, which can be absorbed into the
normalization.  Note ${\rm STr}\Sigma_z \omega |_{
\varphi_{\pm\alpha} = \theta_{\alpha} \pm i\varepsilon} = 0$.

A further simplification results from the fact that $dA$ and the
integration domain for $A$ are Weyl-invariant.  This permits us to
reduce the double sum over ${\hat s}_+ , {\hat s}_-$ to a single sum
over the {\it relative} permutation ${\hat s} := {\hat s}_-^{-1} {\hat
  s}_+^{\vphantom{-1}}$, times an overall factor ${\rm dim} \ {\rm
  W}[{\rm U}(n)\times {\rm U}(n)] = {\rm dim}({\rm S}_n \times {\rm
  S}_n) = n{\rm !}^2$.  Absorbing this constant factor into the
normalization, we obtain for the $n$-level correlation function the
following intermediate result:
        \[
        R_n = {\rm const} \times \sum_{\hat s} (-1)^{|{\hat s}|} 
        \lim_{\varepsilon\to 0} \varepsilon^n \int dA 
        \exp i {\rm STr} A \left( (\hat\theta+i\varepsilon) 
        - {\hat s} (\hat\theta - i\varepsilon) \right) ,
        \]
where $\hat\theta = {\rm diag}(\theta_{\rm B},\theta_{\rm F})$,
$\theta_{\rm B} = \theta_{\rm F} = {\rm diag} (\theta_1 , ... ,
\theta_n)$.  The integral factors into a product of $2n$
one-dimensional exponential integrals over the elements of the
diagonal matrix $A$, $n$ of which are compact $( - 1
\le \cos 2y_i \le 1)$, and the other $n$ are noncompact $(1 \le \cosh
2x_i < \infty)$.  The Weyl group element ${\hat s} =: ({\hat s}_{\rm
B} , {\hat s}_{\rm F}) \in {\rm W}[{\rm U}(n) \times {\rm U}(n)]$ acts
on the BB sector by ${\hat s}_{\rm B}$ and on the FF sector by ${\hat
s}_{\rm F}$.  If $\theta_i = ({\hat s}_{\rm B} \theta)_i$, the
integral over $t_i = \cosh 2x_i$ diverges as $\varepsilon^{-1}$ in the
limit $\varepsilon\to 0$.  The maximal divergence occurs for ${\hat
s}_{\rm B} = {\rm identity}$, in which case a singular factor
$\varepsilon^{-n}$ is produced, canceling the prefactor $\varepsilon^n$ 
and producing a finite result in the limit $\varepsilon \to 0$.
The boundary terms mentioned earlier disappear at this
stage as they contain at most $n-1$ noncompact integrations and
therefore diverge more weakly than $\varepsilon^{-n}$.  What remains
are the $n$ compact integrations over $t_i = \cos 2y_i$.  Using the
elementary integral $\int_{-1}^{+1} dt \ \exp i\pi\theta t = 2 \sin(
\pi\theta) / \pi\theta$ we immediately arrive at the final result:
        \begin{equation}
        \lim_{N\to\infty} R_n(2\pi\theta_1/N , ... , 2\pi\theta_n/N) 
        = \sum_{{\hat s}\in {\rm W}[{\rm U}(n)]}
        (-1)^{|{\hat s}|} \prod_{i=1}^n {\sin \pi(\theta_i - \hat s\theta_i) 
        \over \pi ( \theta_i - \hat s\theta_i )}
        = {\rm Det} \left( {\sin\pi(\theta_i-\theta_j) \over
        \pi(\theta_i-\theta_j)} \right)_{i,j=1,...,n} ,        
        \label{fin_n_level}
        \end{equation}
where the correct value of the normalization constant was restored by
hand.  This result coincides with the one obtained by the Dyson-Mehta 
orthogonal polynomial method\cite{mehta} in the large-$N$ limit.  In 
Sec.~\ref{sec:type_C} we are going to see that the first form of the 
result (\ref{fin_n_level}) extends in a very simple way to the circular
ensemble that is obtained by taking instead of the unitary group the 
symplectic one.  In that sense, this is the ``good'' way of writing 
the result for $R_n$.

\section{Circular Orthogonal Ensemble}
\label{sec:COE}

If an open quantum mechanical system possesses an anti-unitary symmetry
${\cal T}$ (time reversal, for example) and ${\cal T}^2 = + 1$, there
exists a basis of scattering states such that the scattering matrix
$S$ is symmetric: $S = S^{\rm T}$.  The set of symmetric $S$-matrices
can be parameterized in terms of the unitary matrices $U \in {\rm
U}(N)$ by $S = U U^{\rm T}$.  The product $S = U U^{\rm T}$ is
invariant under right multiplication of $U$ by any orthogonal matrix,
which means\cite{dyson70} that $S = U U^{\rm T}$ lives on the coset
space ${\rm U}(N) / {\rm O}(N)$.

The Circular Orthogonal Ensemble (COE) of random-matrix theory is
defined\cite{mehta} by taking $S$ to be distributed according to the 
uniform measure on ${\rm U}(N) / {\rm O}(N)$.  We denote this measure 
by $d\mu(S)$.  The generating function for COE spectral correlators,
$F_{n_+ , n_-}$, is defined as in (\ref{spec_det}) but with the
substitutions ${\rm U}(N) \to {\rm U}(N) / {\rm O}(N)$, $U \to S$, and
$dU \to d\mu(S)$.  Because the invariant measure $d\mu(S)$ is induced
by the Haar measure $dU$ of ${\rm U}(N)$ through the projection ${\rm
U}(N) \to {\rm U}(N) / {\rm O}(N)$, we can write the defining
expression for $F_{n_+ , n_-}$ also as follows:
        \[
        F_{n_+ , n_-}( \{ \theta , \varphi \} ) = \int_{{\rm U}(N)} dU \ 
        \prod_{\alpha=1}^{n_+} {{\rm Det}(1-e^{i\varphi_{+\alpha}}
        UU^{\rm T}) \over {\rm Det}(1 - e^{i\theta_{+\alpha}}UU^{\rm T}) }
        \prod_{\beta=1}^{n_-} {{\rm Det}(1-e^{-i\varphi_{-\beta}}
        {\bar U}U^\dagger) \over {\rm Det}(1 - e^{-i\theta_{-\beta}}
        {\bar U}U^\dagger) } .
        \]
This expression will now be processed by the method of Sec.~\ref{sec:type_A}.
To do so we use the trick of doubling the dimension:
        \[
        {\rm Det} \left( 1 - e^{i\gamma} U U^{\rm T} \right)
        = {\rm Det} \pmatrix{1 &e^{i\gamma}U\cr U^{\rm T} &1\cr} .      
        \]
The extra degree of freedom implied by this doubling will be called 
``quasispin'' and denoted by the symbols $\uparrow, \downarrow$.

After the introduction of quasispin, we express $F_{n_+ , n_-}$
as a Gaussian superintegral in the usual way, see
Sec.~\ref{sec:HS}, Eq.~(\ref{gen_func}).  As before we put
$\{ \omega \} = \{ \theta , \varphi \}$. Then we lump the two terms
containing $U$ into a single one:
        \[
        \bar\psi_{+a\uparrow}^i e^{i\omega_{+a}} U^{ij}
        \psi_{+a\downarrow}^j + \bar\psi_{+a\downarrow}^j 
        (U^{\rm T})^{ji} \psi_{+a\uparrow}^i
        \equiv \phi_{+A}^i U^{ij} \chi_{+A}^j ,
        \]
where the tensors $\phi_+$ and $\chi_+$ have components
        \[
        \phi_{+A}^i = \{ \phi_{+a1}^i , \phi_{+a2}^i \} = \{ 
        \bar\psi_{+a\uparrow}^i , \psi_{+a\uparrow}^i (-1)^{|a|} \} ,
        \quad \chi_{+A}^j = \{ \chi_{+a1}^j , \chi_{+a2}^j \} = \{ 
        e^{i\omega_{+a}} \psi_{+a\downarrow}^j, \bar\psi_{+a\downarrow}^j \}.
        \]
Here $|a| = 0$ if $a = (\alpha,{\rm B})$, and $|a| = 1$ if $a = 
(\alpha,{\rm F})$.  The terms involving $U^\dagger$ are manipulated
in a similar manner:
        \begin{eqnarray}
        &&\bar\psi_{-b\uparrow}^i {\bar U}^{ij} \psi_{-b\downarrow}^j 
        + \bar\psi_{-b\downarrow}^j e^{-i\omega_{-b}} (U^\dagger)^{ji} 
        \psi_{-b\uparrow}^i \equiv \chi_{-B}^j {\bar U}^{ij} \phi_{-B}^i ,
        \quad {\rm where} \nonumber \\
        &&\chi_{-B}^j = \{ \chi_{-b1}^j , \chi_{-b2}^j \} = 
        \{ \bar\psi_{-b\downarrow}^j e^{-i\omega_{-b}}
        , \psi_{-b\downarrow}^j (-1)^{|b|} \} ,
        \quad \phi_{-B}^i = \{ \phi_{-b1}^i , \phi_{-b2}^i \} = 
        \{ \psi_{-b\uparrow}^i , \bar\psi_{-b\uparrow}^i \} .
        \nonumber
        \end{eqnarray}
In the next step we apply the generalized Hubbard-Stratonovitch
transformation (\ref{HS}):
        \[
        \int dU \exp \left( \phi_{+A}^i U^{ij} \chi_{+A}^j +    
        \chi_{-B}^j {\bar U}^{ij} \phi_{-B}^i \right) = \int 
        D\mu_N(Z,\tilde Z) \exp \left( \phi_{+A}^i Z_{AB}^{\vphantom{i}} 
        \phi_{-B}^i + \chi_{-B}^j {\tilde Z}_{BA}^{\vphantom{i}} 
        \chi_{+A}^j \right) .
        \]
Note that the exponential on the right-hand side separates into terms
containing $\phi$ and $\chi$.  The same is true for the terms
$\bar\psi\psi$ that do not involve $U$.  Therefore the integral splits
into two integrals, one over the tensor $\phi$ and the other one over
the tensor $\chi$.  Consider the $\phi$ integral first, and define
an orthogonal matrix $\tau_+$ (``charge conjugation'') by requiring the
tensor $(\tau_+)_{AA'}^{\vphantom{i}}\phi_{+A'}^i$ to have components 
$\{ \psi_{+a\uparrow}^i , \bar\psi_{+a^\uparrow}^i \}$.  The matrix 
that does this job is given in quasispin block decomposition by
        \[
        \tau_+ = \pmatrix{0 &\sigma\cr 1 &0\cr} ,
        \]
where $\sigma$ is the matrix for superparity ($\sigma_{aa'} = (-1)^{|a|}
\delta_{aa'}$).  The same matrix acting in the space of ``negative''
indices (with the appropriate change of dimension) is denoted by
$\tau_-$.  Having introduced these matrices we can write
        \[
        \bar\psi_{+a\uparrow}^i \psi_{+a\uparrow}^i = {\textstyle{1\over 2}}
        \phi_{+A}^i (\tau_+^{\vphantom{i}})_{AA'}^{\vphantom{i}} \phi_{+A'}^i 
        , \quad {\rm and} \quad
        \bar\psi_{-b\uparrow}^i \psi_{-b\uparrow}^i = {\textstyle{1\over 2}}
        \phi_{-B}^i (\tau_-^{\vphantom{i}})_{BB'}^{\vphantom{i}} \phi_{-B'}^i .
        \]
On collecting terms we encounter the integral
        \begin{eqnarray}
        &&\int D\phi \ \exp \left( \phi_{+A}^i Z_{AB}^{\vphantom{i}} 
        \phi_{-B}^i -  {\textstyle{1\over 2}} \phi_{+A}^i 
        (\tau_+^{\vphantom{i}})_{AA'}^{\vphantom{i}} \phi_{+A'}^i - 
        {\textstyle{1\over 2}} \phi_{-B}^i (\tau_-^{\vphantom{i}})_{BB'}^
        {\vphantom{i}} \phi_{-B'}^i \right)
        \nonumber \\
        &=& (-1)^{nN/2} {\rm SDet}^{-N/2} \pmatrix{\tau_+ &-Z\cr 
        -Z^{\rm T}\sigma &\tau_-\cr}
        = {\rm SDet}^{-N/2} \left( 1 - Z^{\rm T} \tau_+^{\vphantom{-1}} 
        Z \tau_-^{-1} \right) .
        \nonumber
        \end{eqnarray}
Here $(Z^{\rm T})_{BA} = Z_{AB} (-1)^{(|A|+1)|B|}$ is the
supertranspose, with $|A| = 0$ if $A = +(\alpha,{\rm B})\uparrow$,
and $|A| = 1$ if $A = +(\alpha,{\rm F})\uparrow$.  The
$\chi$ integral is done in a similar fashion and gives ${\rm
SDet}^{-N/2} \left( 1 - \tilde Z \tau_+^{-1} e^{i\omega_+}
\tilde Z^{\rm T} e^{-i\omega_-} \tau_-^{\vphantom{-1}} \right)$.  
By making the variable substitution
        \[
        Z \mapsto e^{i\omega_+ / 4} iZ e^{-i\omega_- / 4} , \quad
        \tilde Z \mapsto - e^{i\omega_- / 4} i\tilde Z e^{i\omega_+ / 4} ,
        \]
we bring the final result into the symmetrical form
        \begin{eqnarray}
        F_{n_+ , n_-}( \{ \omega \} ) = \int D\mu_N(Z,\tilde Z) \ 
        &&{\rm SDet}^{-N/2} \left( 1 + Z^{\rm T} \tau_+^{\vphantom{-1}} 
        e^{i\omega_+ / 2} Z e^{-i\omega_- / 2} \tau_-^{-1} \right)
        \nonumber \\ 
        \times &&{\rm SDet}^{-N/2} \left( 1 + \tilde Z \tau_+^{-1} 
        e^{i\omega_+ / 2} {\tilde Z}^{\rm T} e^{-i\omega_- / 2}
        \tau_-^{\vphantom{-1}} \right) ,
        \nonumber
        \end{eqnarray}
which is exact for all $N$.  The matrices $Z, \tilde Z$
parameterize the complex coset space ${\rm Gl}(2n|2n)/$ ${\rm Gl}(2n_+|
2n_+)\times{\rm Gl}(2n_-|2n_-)$.

We turn to the large-$N$ limit.  In order for this limit to be
nontrivial, we must again take the difference $\omega_+ - \omega_-$ to be 
of order 1/N.  By $D\mu_N(Z,\tilde Z) = D(Z,\tilde Z) \exp N
{\rm STr} \ln (1-\tilde Z Z)$ the integrand at $\omega_+ = \omega _-
= {\rm const} \times {\rm identity}$ can be written as
        \[
        \exp \left( N {\rm STr} \ln (1 - \tilde Z Z) - {N\over 2}
        {\rm STr} \ln (1 + Z^{\rm T} \tau_+^{\vphantom{-1}} Z\tau_-^{-1}) 
        - {N\over 2} {\rm STr} \ln (1 + \tilde Z \tau_+^{-1} 
        \tilde Z^{\rm T} \tau_-^{\vphantom{-1}}) \right) .
        \]
For large $N$, the integral is dominated by contributions from the
subspace determined by the equation
        \[
        Z = - \tau_+^{-1} {\tilde Z}^{\rm T} \tau_-^{\vphantom{-1}} .
        \]
This subspace is referred to as the saddle-point manifold.
Integration over the Gaussian fluctuations normal to the saddle-point
manifold gives unity by supersymmetry, in the large-$N$ limit. 

By restricting the integrand to the saddle-point manifold, expanding
with respect to $\omega_+ - \omega _-$ and keeping only the terms
that survive for $N \to \infty$, we obtain
        \[
        \lim_{N\to\infty} F_{n_+ , n_-} ( 2\pi\{ \omega \} /N )
        = \int D(Z,\tilde Z) \exp i\pi{\rm STr} 
        \left( \omega_+ Z\tilde Z (1-Z \tilde Z)^{-1} - \omega_- 
        \tilde Z Z(1-\tilde Z Z)^{-1} \right) .
        \nonumber
        \]
This has the same form as for the CUE, except for the replacement $N
\to N/2$, the doubling of the dimensions of the supermatrices $Z,
\tilde Z$, and the imposition of the constraint $Z = - \tau_+^{-1}
{\tilde Z}^{\rm T} \tau_-^{\vphantom{-1}}$.  On putting $\omega 
\equiv \{ \omega \} = {\rm diag}(\omega_+ , \omega_-)$ we can 
re-express the result in the form
        \[
        \lim_{N\to\infty} F_{n_+ , n_-} ( 2\pi\omega/N )
        = \int_{M_{\rm B}\times M_{\rm F}} Dg_H \exp 
        i\pi {\rm STr} \omega ( g \Sigma_z g^{-1} - \Sigma_z ) / 2 .
        \]
The relation between $Q = g \Sigma_z g^{-1}$ and $Z,\tilde Z$ is 
formally the same as before.  The constraint $Z = - \tau_+^{-1} 
{\tilde Z}^{\rm T} \tau_-^{\vphantom{-1}}$ is
known\cite{vwz,mrz_suprev} to reduce the complex superspace ${\rm
Gl}(2n|2n) / {\rm Gl}(2n_+|2n_+) \times{\rm Gl}(2n_-|2n_-)$ to the
complex submanifold ${\rm Osp}(2n|2n) / {\rm Osp}(2n_+|2n_+) \times
{\rm Osp}(2n_-|2n_-)$.  The conjugation relations $\tilde Z_{\rm
BB}^{\vphantom{\dagger}} = Z_{\rm BB}^\dagger$ and $\tilde Z_{\rm
FF}^{\vphantom{\dagger}} = - Z_{\rm FF}^\dagger$ translate into
        \[
        M_{\rm B} = {\rm SO}(2n_+ , 2n_-) / {\rm SO}(2n_+) \times
        {\rm SO}(2n_-) , \quad
        M_{\rm F} = {\rm Sp}(2n_+ + 2n_-) / {\rm Sp}(2n_+) \times
        {\rm Sp}(2n_-) .
        \]
For $n_+ = n_- = 1$, the above expression for the generating function
$F_{n_+ , n_-}$ can be calculated by introducing Efetov's polar coordinates.  
The resulting formula for the two-level correlation function $R_2$ coincides 
with that of\cite{efetov}.  It is not clear at present how to do this
calculation for general $n_+ , n_-$.  [Harish-Chandra's formula does
{\it not} extend to this case as neither $\Sigma_z$ nor $\omega =
{\rm diag}(\omega_+ , \omega_-)$ are elements of ${\rm Lie}\left(
{\rm Osp}(2n|2n)\right)$.]

\section{Circular Ensemble of type C}
\label{sec:type_C}

Dyson's Circular Ensembles COE and CSE are constructed by starting
from the unitary group and passing to the coset spaces ${\rm U}(N) /
{\rm O}(N)$ and ${\rm U}(2N)/{\rm Sp}(2N)$.  Other circular ensembles
of relevance\cite{az_prb} to mesoscopic physics can be obtained by
taking for the starting point the orthogonal or symplectic group
instead of the unitary one.  Again the option of forming various
coset spaces exists.  Of these possibilities the technically simplest
one is to consider an ensemble of scattering matrices $S$ drawn at
random from the symplectic group ${\rm Sp}(2N)$, with no coset
projection done.  This is what we shall do in the present section.
The probability distribution for $S$ will be taken to be the Haar
measure $dS$ of ${\rm Sp}(2N)$.  The circular ensemble so defined is
called ``type $C$''.  A physical system where such an ensemble can be
realized is a chaotic Andreev quantum dot\cite{az_prl} with
time-reversal symmetry broken by a weak magnetic field.

The defining equations of ${\rm Sp}(2N)$ are
        \[
        {S^{-1}}^\dagger = S = {\cal C} {S^{-1}}^{\rm T} {\cal C}^{-1} ,
        \quad {\rm where} \quad {\cal C} = i\sigma_y \otimes 1_N =
        \pmatrix{0 &1_N\cr -1_N &0\cr} 
        \]
is the symplectic unit in $2N$ dimensions.  The polar (or Cartan)
decomposition of an element $S \in {\rm Sp}(2N)$ has the form 
$S = k e^{i\hat\theta} k^{-1}$ where $\hat\theta = \sigma_z
\otimes {\rm diag}(\theta_1, ..., \theta_N)$ and $\sigma_z = 
{\rm diag}(+1,-1)$.  Thus if $\theta_1$ is an eigenphase of $S$,
then so is $-\theta_1$.  This symmetry under reflection $\theta
\to -\theta$ is called a ``particle-hole'' symmetry.  

All information about the eigenphase correlations of the circular
random-matrix ensemble of type $C$ is contained in the generating function
        \[
        F_n(\{ \theta , \varphi \}) = \int_{{\rm Sp}(2N)} dS \ 
        \prod_{\alpha=1}^{n} {{\rm Det}(1-e^{i\varphi_{\alpha}}S) \over
        {\rm Det}(1 - e^{i\theta_{\alpha}}S) }  \qquad \quad
        ({\rm Im}\theta_{\alpha} > 0) .
        \]
This expression is simpler than the corresponding one for the CUE
because ${\rm Det}(1-zS) = {\rm Det}(1-zS^\dagger)$ as a result
of the particle-hole symmetry.
By putting $\{ \omega_a \} = \{ \theta_\alpha , \varphi_\alpha \}$
and proceeding as in Sec.~\ref{sec:HS}, we get a 
Gaussian superintegral representation for $F_n$:
        \[
        F_n(\{ \omega \}) = \int D(\psi,\bar\psi) \int_{{\rm Sp}(2N)} dS
        \ \exp - \bar\psi_{a}^i (\delta^{ij} - e^{i\omega_{a}}
        U^{ij}) \psi_{a}^j .
        \]
The components $\psi_a^i, \bar\psi_a^i$ of the tensors $\psi, \bar\psi$ 
are bosonic for $a = (\alpha,{\rm B})$ and fermionic for $a = (\alpha,
{\rm F})$.  The analog of (\ref{HS}) reads:
        \[
        \int_{{\rm Sp}(2N)} dS \exp
        \bar\psi_{a}^i S^{ij} \psi_{a}^j
        = \int_{M_{\rm B}\times M_{\rm F}} D\mu_N(Z,\tilde Z) 
        \exp \left( {\textstyle{1\over 2}}
        \bar\psi_{a}^i Z_{ab}^{\vphantom{i}} {\cal C}^{ij} \bar\psi_{b}^j 
        - {\textstyle{1\over 2}} \psi_{a}^i (-1)^{|a|} 
        {\tilde Z}_{ab}^{\vphantom{i}} {\cal C}^{ij}\psi_{b}^j 
        \right) .
        \]
As before, $D\mu_N(Z,\tilde Z) = D(Z,\tilde Z) \ {\rm SDet}(1 - \tilde Z
Z)^N$, but $Z$ and $\tilde Z$ are now {\it square} supermatrices and are
subject to the symmetry conditions
        \[
        Z_{ab} = (-1)^{|a| + |b| + |a||b| + 1} Z_{ba}, \quad
        \tilde Z_{ab} = (-1)^{|a||b| + 1} \tilde Z_{ba} .
        \]
These conditions can be succinctly written as
        \[
        Z = - Z^{\rm T} \sigma , \quad
        \tilde Z = - \sigma \tilde Z^{\rm T} ,
        \]
where $\sigma$ is the superparity as before.  They express 
the fact\cite{mrz_suprev} that $Z, \tilde Z$ parameterize the complex 
supermanifold $G / H = {\rm Osp}(2n|2n) / {\rm Gl}(n|n)$, see also below.
The integration domain $M_{\rm B} \times M_{\rm F}$ is fixed by
        \[
        \tilde Z_{\rm BB}^{\vphantom{\dagger}} = Z_{\rm BB}^\dagger, \quad 
        \tilde Z_{\rm FF}^{\vphantom{\dagger}} = - Z_{\rm FF}^\dagger ,
        \]
and the requirement that all eigenvalues of the positive hermitian
$n\times n$ matrix $\tilde Z_{\rm BB} Z_{\rm BB}$ be less than unity.  
It can be shown\cite{mrz_suprev} that this means
        \[
        M_{\rm B} = {\rm SO}^*(2n) / {\rm U}(n),
        \quad
        M_{\rm F} = {\rm Sp}(2n) / {\rm U}(n).
        \]
$(G / H , M_{\rm B} \times M_{\rm F})$ is called a Riemannian symmetric 
superspace of type $D{\rm III}|C{\rm I}$.

In the simplest case $n = 1$
the supermatrices $Z, \tilde Z$ have the form
        \[
        Z = \pmatrix{0 &\zeta_1 \cr \zeta_1 &z_1\cr} , \quad
        \tilde Z = \pmatrix{0 &\zeta_2 \cr -\zeta_2 &z_2\cr} ,
        \]
where $z_1, z_2$ are complex commuting and $\zeta_1 , \zeta_2$ are
complex anticommuting numbers.  It is seen that the BB sector $M_{\rm B}$
is trivial in this case (in fact ${\rm SO}^*(2)/{\rm U}(1)$ consists of
just a single point).  The variable $z_2 = - \bar z_1$ of the
FF sector can be interpreted as the complex stereographic coordinate of
a two-sphere ${\rm S}^2 \simeq {\rm Sp}(2) / {\rm U}(1)$.  It is
very important to note\cite{mrz_suprev} that the variables $\zeta_1 ,
\zeta_2$ {\it are not related by any kind of complex conjugation}.

The proof of the generalized Hubbard-Stratonovitch identity for the
present case is closely analogous to that for the CUE, presented in
detail in the Appendix.  (To understand the following remark you must
study that appendix first.) The key idea is to consider the
generalized coherent states
        \begin{eqnarray}
        | Z \rangle &:=& \exp \left( {\textstyle{1\over 2}}
        {\bar c}_{a}^i Z_{ab}^{\vphantom{i}} {\cal C}^{ij} 
        {\bar c}_{b}^j \right) 
        | 0 \rangle \ {\rm SDet}(1-\tilde Z Z)^{N/2} ,
        \nonumber \\
        \langle Z | &:=& {\rm SDet}(1-\tilde Z Z)^{N/2} \ \langle 0 | 
        \exp \left( {\textstyle{1\over 2}} c_{a}^i (-1)^{|a|} 
        \tilde Z_{ab}^{\vphantom{i}} {\cal C}^{ij} c_{b}^j \right) ,
        \nonumber
        \end{eqnarray}
built from an absolute vacuum $c_a^i |0\rangle = 0$ by repeatedly
acting with Bose and Fermi creation and annihilation operators obeying
the supercommutation relations $[ c_a^i , {\bar c}_b^j ] = \delta^{ij}
\delta_{ab}$, and to exploit the fact that $P = \int D(Z,\tilde Z)
|Z\rangle\langle Z|$ projects on the singlet sector of the symplectic
group acting on Fock space by ${\bar c}_a^i \mapsto S^{ij} {\bar
c}_a^j$.

Use of the generalized Hubbard-Stratonovitch transformation leads to
        \begin{eqnarray}
        F_n(\{ \omega \}) &=& \int D\mu_N(Z,\tilde Z) \ {\rm SDet}^{-N}
        \pmatrix{1 &-e^{i\omega} Z e^{i\omega}\cr -\tilde Z &1\cr}
        \nonumber \\
        &=& \int D(Z,\tilde Z) \ {\rm SDet}^{-N} \left( 1 - (1-\tilde 
        Z Z)^{-1} \tilde Z ( e^{i\omega} Z e^{i\omega} - Z ) \right) ,
        \nonumber
        \end{eqnarray}
where $\omega$ stands for the diagonal matrix with elements $\omega_a$.
To arrive at an equivalent $Q$-integral representation, we let $g$ run 
through the orthosymplectic Lie supergroup $G = {\rm Osp}(2n|2n)$ defined 
by
        \[
        g = \pmatrix{A &B\cr C &D\cr} = \pmatrix{0 &1 \cr
        \sigma &0\cr} \pmatrix{A^{\rm T} &C^{\rm T}\cr B^{\rm T}     
        &D^{\rm T}\cr}^{-1} \pmatrix{0 &\sigma\cr 1 &0\cr} ,
        \]
and set $Z = BD^{-1}$, $\tilde Z = CA^{-1}$,
        \[
        Q := g \Sigma_z g^{-1} := 
        \pmatrix{1 &Z\cr \tilde Z &1\cr}
        \pmatrix{1 &0\cr 0 &-1\cr}
        \pmatrix{1 &Z\cr \tilde Z &1\cr}^{-1}
        =: \pmatrix{Q_{++} &Q_{+-}\cr Q_{-+} &Q_{--}\cr} ,
        \]
as before.  The subgroup $H \subset G$ of elements $h = {\rm diag}(A,
{A^{-1}}^{\rm T})$ satisfying $h\Sigma_z h^{-1} = \Sigma_z$ is
isomorphic to ${\rm Gl}(n|n)$.  Hence the elements of the supermatrix
$Q$ are functions on the coset space $G / H = {\rm Osp}(2n|2n) / {\rm
Gl}(n|n)$.  By simple manipulations we obtain
        \[
        F_{n}(\{ \omega \}) = \int DQ \ {\rm SDet}^{-N} \left(
        1 - Q_{--} + Q_{-+} e^{i\omega} (1+Q_{++})^{-1} Q_{+-}
        e^{i\omega} \right) ,
        \]
which is to be compared with (\ref{Q_integral}).

In the large-$N$ limit the exact expression for $F_n$ simplifies to
        \[
        \lim_{N\to\infty} F_n(\pi \{ \omega \} / N) = \int_{M_{\rm B}
        \times M_{\rm F}} Dg_H  \exp i\pi {\rm STr} 
        \hat\omega ( g \Sigma_z g^{-1} - \Sigma_z ) / 2 ,
        \]
where $\hat\omega = {\rm diag}(\omega , -\omega)$.  Note that this is
formally identical to (\ref{fin_res_g}).  The CUE $n$-level
correlation function $R_n$ was extracted from that equation in
Sec.~\ref{sec:largeN}.  It turns out that the corresponding
calculation can be done in the present case, too, by extending
Harish-Chandra's formula (\ref{harish-chandra}) to the orthosymplectic
Lie supergroup.  [In contrast with the situation for the COE, the
matrices $\hat\omega$ and $\Sigma_z$ now {\it are} elements of ${\rm
Lie}\left({\rm Osp}(2n|2n)\right)$.]  To keep the present paper within 
size, let me omit the details of this calculation (which in any case is 
very similar to that of Sec.~\ref{sec:largeN}) and simply state the 
result for $R_n$ thus obtained:
        \[
        \lim_{N\to\infty} R_n(\pi\theta_1/N , ... , \pi\theta_n/N) = 
        \sum_{{\hat s}\in {\rm W}[{\rm Sp}(2n)]}
        (-1)^{|{\hat s}|} \prod_{i=1}^n {\sin \pi(\theta_i - \hat s\theta_i) 
        \over \pi ( \theta_i - \hat s\theta_i )} .
        \]
The Weyl group ${\rm W}[{\rm Sp}(2n)]$ is generated by the operations
of transposition $\theta_{i-1} \leftrightarrow \theta_{i}$ $(i = 2, ..., 
n)$ and reflection $\theta_1 \mapsto -\theta_1$, all of which have
odd parity $|{\hat s}| = 1$.  Please observe the perfect analogy to
(\ref{fin_n_level}): the Weyl group of ${\rm U}(n)$ has simply been
replaced by the Weyl group of ${\rm Sp}(2n)$.  Specialization to 
$n = 1, 2$ gives
        \begin{eqnarray}
        \lim_{N\to\infty} R_1 (\pi\theta_1 / N) &=& 1 - {\sin 2\pi\theta_1
        \over 2\pi\theta_1} ,
        \nonumber \\
        \lim_{N\to\infty} R_2 (\pi\theta_1 / N , \pi\theta_2 / N) &=& 
        \left( 1 - {\sin 2\pi\theta_1 \over 2\pi\theta_1} \right)
        \left( 1 - {\sin 2\pi\theta_2 \over 2\pi\theta_2} \right)
        \nonumber \\
        &&- \left( {\sin\pi(\theta_1-\theta_2) \over \pi(\theta_1-\theta_2)} 
        - {\sin\pi(\theta_1 +\theta_2) \over \pi(\theta_1+\theta_2)} 
        \right)^2 .
        \nonumber 
        \end{eqnarray} 
Precisely the same expressions were obtained for the correlation
functions of the {\it Gaussian} random-matrix ensemble of type $C$
in\cite{az_prb}.  The reasoning used there was heuristic and took
recourse to the mapping on a one-dimensional free Fermi gas with 
Dirichlet boundary conditions at the origin $\theta = 0$.  This 
coincidence of results is of course expected from the large-$N$ 
equivalence of the Circular and Gaussian Ensembles.

\section{Summary and Outlook}
\label{sec:conclusion}

The key result of this paper is the Hubbard-Stratonovitch identity
(\ref{HS}), relating an integral over the unitary group ${\rm U}(N)$
to an integral over a Riemannian symmetric superspace of type $A{\rm
  III}|A{\rm III}$ (Efetov's unitary $\sigma$-model space).  The
detailed proof given in the Appendix derives this identity from the
standard properties of generalized coherent states.  Instead of ${\rm
  U}(N)$ one can also consider the symplectic group ${\rm Sp}(2N)$, or
the orthogonal group in an even number of dimensions, ${\rm SO}(2N)$.
In these cases, identities similar to (\ref{HS}) exist and relate the
corresponding group integrals to integrals over Riemannian symmetric
superspaces of type $D{\rm III}|C{\rm I}$ and $C{\rm I}|D{\rm III}$
respectively. For ${\rm Sp}(2N)$ this was described briefly in
Sec.~\ref{sec:type_C}.  The case of ${\rm SO}(2N)$ was not discussed
here and is left for future work.

In the present paper the generalized Hubbard-Stratonovitch
transformation was applied to Dyson's Circular Unitary Ensemble (CUE,
or type $A$) and the circular ensemble of type $C$.  In both cases the
large-$N$ limit of the $n$-level correlation function was calculated
for all $n$, by appropriate extensions of Harish-Chandra's formula.
The method is not restricted to spectral correlations but can be used
for wave amplitude correlations and transport coefficients as well.
It can also be adapted to other symmetry classes.  The trick that
works for Dyson's Circular Orthogonal Ensemble (COE, or type $A{\rm
  I}$) is to write the elements $S$ of the COE in the form $S = U
U^{\rm T}$ with $U \in {\rm CUE}$ and then proceed as in the unitary
case.  A similar trick works for Dyson's Circular Symplectic Ensemble
(type $A{\rm II}$), the Circular Ensemble of type $C{\rm I}$ and, in
fact, any one of the large class\cite{mrz_suprev} of circular
ensembles.

As was mentioned in the introduction, nontrivial applications to
random network models are expected.  What puts these within reach is
the fact that the Hubbard-Stratonovitch scheme (\ref{HS}) is valid for
all $N$, including $N = 1$.  This permits to transform a network model
with random ${\rm U}(1)$ phases into a theory of coupled supermatrices
$Z, \tilde Z$.  Doing so for the Chalker-Coddington model\cite{lc},
for example, and taking a continuum limit, one obtains\cite{mrz_cc}
the two-dimensional nonlinear $\sigma$ model augmented by Pruisken's
topological term.

Another realm of fruitful application of the identity (\ref{HS}) will
be quantum chaotic maps.  Andreev et al.\cite{aasa} have recently
argued that energy averaging for deterministic Hamiltonian systems can
be used to derive a ``ballistic''\cite{mk} nonlinear $\sigma$ model,
in much the same way as impurity averaging for disordered systems
leads to the usual ``diffusive'' $\sigma$ model.  When one goes from
Hamiltonian systems to periodically driven ones or maps, the role of
the Hamiltonian passes to a unitary operator $U$, the Floquet or
time-evolution operator.  As the spectrum of $U$ lies on the unit
circle in ${\Bbb C}$, an approach analogous to that of Andreev et al.
will employ ${\rm U}(1)$-phase averaging instead of energy averaging.
It is clear that, by carrying out this phase average with the help of
the identity (\ref{HS}), one will be able to derive a nonlinear
$\sigma$ model for maps.  Work in this direction is in
progress\cite{aaz}.

Finally, let me mention that the identity (\ref{HS}) is not restricted
to the supersymmetric case.  Similar formulas can also be derived
rigorously when the spinor $\psi$ is purely bosonic or fermionic.
(In the bosonic case, convergence of all integrals places a lower
bound on the allowed values of $N$.)  The fermionic version of (\ref{HS})
is particularly exciting as it promises nontrivial applications to 
lattice gauge theory, by enabling an exact transformation from the
gauge degrees of freedom to meson fields.  (Note that the usual
plaquette term ${\rm Tr}UUUU$ of lattice gauge theory can be generated
by coupling a number of heavy fermionic ghosts to the gauge field
via $\bar\psi U \psi$.)  In this way, one may succeed in constructing a
nonperturbative proof of a result due to Witten\cite{witten}, who
summed planar diagrams to argue that large-$N$ quantum chromodynamics
is equivalent to a weakly coupled theory of mesons, in the low-energy
limit.  Sadly, although the exact gluon-meson transformation works
for the gauge groups ${\rm U}(N)$ and ${\rm Sp}(2N)$ (and, presumably,
for ${\rm SO}(2N)$ as well), it does not easily extend to ${\rm SU}(N)$,
except for $N = 2$, where one has the accidental isomorphism
${\rm SU}(2) \simeq {\rm Sp}(2)$.

This work was supported in part by the National Science Foundation
under Grant No. PHY94-07194, and by the Deutsche Forschungsgemeinschaft,
SFB 341, K\"oln-Aachen-J\"ulich. 

\section{Appendix}

Given a compact Lie group $K$ and a unitary irreducible action $T$ of
$K$ on a module $V$ with lowest (or highest) weight $| 0 \rangle$, one
can consider objects of the form $T_k | 0 \rangle$ $(k \in K)$.  These
are called {\it generalized coherent states}, and they have many nice
properties\cite{gcs,onofri,perelomov}.  Most importantly, they resolve
the unit operator on $V$:
        \[
        1_V = \int_K dk \ T_k^{\vphantom{\dagger}} 
        | 0 \rangle \langle 0 | T_k^\dagger ,
        \]
where $dk$ denotes the Haar measure of $K$, normalized by $\int_K dk
\big| \langle 0 | T_k | 0 \rangle \big|^2 = 1$.  A well-known
application of this is the derivation of coherent-state path integrals
for quantum spin systems, in which case $K = {\rm SU}(2)$ and $V$ is a
spin-$S$ representation space of ${\rm SU}(2)$.  The objective of the
present appendix is to extend some of the mathematics of generalized
coherent states to the case where instead of $K$ we have the
supergroup $G = {\rm Gl}(n|n)$ that acts on Efetov's complexified
coset space $G / H$ of unitary symmetry.  This will eventually yield 
an easy proof of the identity (\ref{HS}).

We start by defining a group action of ${\rm Gl}(n|n)$ on a Fock space
of bosons and fermions and developing various useful structures that
come with it.  The general context and our notations have been laid
down in Secs.~\ref{sec:HS}, \ref{sec:geometry}.  Recall that the
tensor $\psi$ has components $\psi_{+a}^i$, $\psi_{-b}^i$, where $i =
1, ..., N$, and $a = (\alpha,\sigma)$, $b = (\beta,\sigma)$ are
composite indices with range $\alpha = 1, ..., n_+$; $\beta = 1,
...,n_-$; and $\sigma = {\rm B,F}$ (Bosons and Fermions).  For
notational convenience, we introduce a composite index $A$ taking
values $+a$ or $-b$.  Thus, $\{ \psi_A^i \} = \{ \psi_{+a}^i ,
\psi_{-b}^i \} $.  The terminology we use is to refer to the upper
index $i$ as ``color'' and the lower index $A$ as ``flavor''.  Now we
introduce creation and annihilation operators ${\bar c}_A^i, c_A^i$
for bosons and fermions, which obey the canonical supercommutation
relations
        \[
        [ c_A^i , \bar c_B^j ] := c_A^i \bar c_B^j - 
        (-1)^{|A||B|} \bar c_B^j c_A^i = \delta_{AB} \delta^{ij} ,
        \]
where $|A| = 0$ if $A = \pm(\alpha,{\rm B})$ and $|A| = 1$ if $A =
\pm(\alpha,{\rm F})$.  These operators act on a Fock space with vacuum
$| 0 \rangle$ defined by $c_A^i |0\rangle = 0$.  They can be viewed as
the quantum mechanical counterparts of the ``classical'' variables
$\bar\psi_A^i , \psi_A^i$.

For the following it is very convenient (though not necessary) to make 
the transformation
        \[
        c_{-b}^i \mapsto \bar c_{-b}^i , \quad 
        \bar c_{-b}^i \mapsto (-1)^{|b|+1} c_{-b}^i ,
        \]
which is canonical, i.e. leaves the supercommutation relations
unchanged.  (The operators $c_{+a}^i$, ${\bar c}_{+a}^i$ are not
transformed.)  If the Fock space is endowed with its usual inner
product (so that ${\bar c}_A^i$ is adjoint to $c_A^i$), the 
transformation is {\it nonunitary} in the bosonic sector since, 
if $b , b^\dagger$ are adjoints of each other, the transformed
canonical pair $b^\dagger , -b$ no longer are.  One can check that 
{\it no mathematical problems are caused by this nonunitarity}.  
After the transformation, the Fock space vacuum $|0\rangle$ and 
its dual $\langle 0 |$ are annihilated by
        \[
        c_{+a}^i | 0 \rangle = \bar c_{-b}^j | 0 \rangle = 0 
        \quad {\rm and} \quad
        \langle 0 | \bar c_{+a}^i = \langle 0 | c_{-b}^j = 0 .
        \]
We may imagine that the ``positive-energy states'' $(+a)$ are empty
while the ``negative-energy states'' $(-b)$ have been filled.  Of course,
for unitary bosons no such thing as a filled negative-energy sea exists,
which is another way of seeing why the above canonical transformation  
must be nonunitary.  Note also
        \[
        \sum_{\beta=1}^{n_-} {\bar c}_{-\beta,{\rm F}}^i c_{-\beta,{\rm F}}^j 
        |0\rangle = |0\rangle \times (+ \delta^{ij} n_- ) , \qquad
        \sum_{\beta=1}^{n_-} {\bar c}_{-\beta,{\rm B}}^i c_{-\beta,{\rm B}}^j 
        |0\rangle = |0\rangle \times (- \delta^{ij} n_-) .
        \]
(The summation convention has been temporarily suspended for clarity.)

Consider now the set of operators $E_{AB}^{ij} := {\bar c}_A^i c_B^j$.
By the commutation relations for the $c, \bar c$ these bilinears form
a ${\rm gl}(nN,nN)$ Lie superalgebra $(n = n_+ + n_-)$:\footnote{One
advantage gained by the nonunitary canonical transformation $c_{-b}^i
\mapsto \bar c_{-b}^i$, $\bar c_{-b}^i \mapsto (-1)^{|b|+1} c_{-b}^i$
is that it allows us to present all ${\rm gl}(nN,nN)$ generators in
the form $\bar c c$, whereas in the original presentation they would
have been be of various types $\bar c c$, $\bar c \bar c$ and $c c$.}
        \begin{eqnarray}
        [ E_{AB}^{ij} , E_{CD}^{kl} ] &:=& 
        E_{AB}^{ij} E_{CD}^{kl} - (-1)^{(|A|+|B|)(|C|+|D|)}
        E_{CD}^{kl} E_{AB}^{ij} \nonumber \\
        &=& \delta^{jk} \delta_{BC} E_{AD}^{il} - (-1)^{(|A|+|B|)
        (|C|+|D|)}\delta^{li} \delta_{DA} E_{CB}^{kj} .
        \nonumber
        \end{eqnarray}  
Two subalgebras we shall have use for are the ${\rm gl}(N)$ Lie
algebra generated by the flavor-singlet operators $\sum_A {\bar c}_A^i
c_A^j$, and the ${\rm gl}(n,n)$ Lie superalgebra generated by the
color-singlet operators $\sum_i {\bar c}_A^i c_B^i$.  These
subalgebras commute and, moreover, they are {\it maximal} relative to
each other, i.e. the maximal subalgebra of ${\rm gl}(nN,nN)$ whose
elements commute with all elements of ${\rm gl}(N)$, is ${\rm
gl}(n,n)$, and vice versa.

In the sequel, we focus attention on that part of Fock
space where all ${\rm gl}(N)$ generators vanish identically:
        \[
        \sum_A {\bar c}_A^i c_A^j |{\rm flavor} \ 
        {\rm state}\rangle \equiv 0 .
        \]
This subspace contains the vacuum:
        \[
        \sum_A {\bar c}_A^i c_A^j |0\rangle =
        \sum_{\beta = 1}^{n_-} \left( {\bar c}_{-\beta,{\rm B}}^i
        c_{-\beta,{\rm B}}^j + {\bar c}_{-\beta,{\rm F}}^i 
        c_{-\beta,{\rm F}}^j \right) |0\rangle = |0\rangle \times
        \delta^{ij} ( - n_- + n_- ) = 0 ,
        \]
and will be called the ``flavor sector'' (or color-neutral sector).  We
claim that the action of ${\rm gl}(n,n)$ on the flavor sector is {\it
irreducible}, i.e. every state of the flavor sector can be reached by a
multiple action of the flavor operators $\sum_i {\bar c}_A^i c_B^i$ on
the vacuum.  The argument goes as follows.  By employing the
occupation number representation of Fock space, one easily sees that
${\rm gl}(nN,nN)$ acts irreducibly on the subspace of Fock space
selected by the condition $\sum_{A,i} {\bar c}_A^i c_A^i = 0$, which
contains the color-neutral sector and might be called the
``charge-neutral'' sector.  Thus every charge-neutral state $|{\cal
N}\rangle$ can be obtained by acting on the vacuum (or lowest-weight)
state $|0\rangle$ with the raising operators, ${\bar c}_{+a}^i
c_{-b}^j$, of ${\rm gl}(nN,nN)$:
        \[
        |{\cal N}\rangle = \sum F_{a_1 b_1 ... a_r b_r}^{i_1 j_1 ... i_r j_r}
        {\bar c}_{+a_1}^{i_1} c_{-b_1}^{j_1} ...        
        {\bar c}_{+a_r}^{i_r} c_{-b_r}^{j_r} | 0 \rangle .
        \]
If $R \in {\rm Gl}(N)$ is a rotation in color space, such a state
transforms as a number of copies of the vector representation ${\bar
c}_A^i \mapsto \sum_j R^{ij} {\bar c}_A^j$ and the same number of
copies of the co-vector representation $c_A^i \mapsto \sum_j
(R^{-1})^{ji} c_A^j$.  Now assume $|{\cal N}\rangle$ to be color-neutral,
i.e. $\sum_A {\bar c}_A^i c_A^j |{\cal N}\rangle = 0$.  In order for this
equation to hold, the indices of the expansion coefficients have to be
contracted pairwise ($\pi$ here stands for permutations):
        \[
        F_{a_1 b_1 ... a_r b_r}^{i_1 j_1 ... i_r j_r}
        = \sum_{\pi \in {\rm S}_r} f(a_1 b_1 ... a_r b_r ; \pi)
        \delta_{i_1 \pi(j_1)} ... \delta_{i_r \pi(j_r)} ,
        \]
by a standard result on the reduction of tensor-product representations
of ${\rm Gl}(N)$.  This form of the expansion coefficients permits to 
express the operators that create $|{\cal N}\rangle$ out of the vacuum,
in terms of the raising operators $\sum_i {\bar c}_{+a}^i c_{-b}^i$ of 
the flavor superalgebra ${\rm gl}(n,n)$.  Thus the vacuum $| 0 \rangle$
is a cyclic vector for the action of ${\rm gl}(n,n)$ on the flavor
sector.  Because the action of every particle-hole creation operator
$\sum_i {\bar c}_{+a}^i c_{-b}^i$ can be undone by the corresponding 
annihilation operator $\sum_i {\bar c}_{-b}^i c_{+a}^i$, the existence
of a cyclic vector implies irreducibility. (q.e.d.)

In addition to the Lie superalgebra ${\rm gl}(n,n)$ we shall need a
corresponding Lie supergroup, ${\rm Gl}(n|n)$.  Recall that $G := {\rm
Gl}(n|n)$ is defined as the group of regular complex supermatrices of
dimension $(n+n)\times (n+n)$.\footnote{The change in notation from
the Lie superalgebra ${\rm gl}(n,n)$ (komma) to the Lie supergroup
${\rm Gl}(n|n)$ (vertical bar) reminds us of the fact that the
definition of the latter requires the specification of some parameter
Grassmann algebra, whereas the former does not.}  We associate with
each element $g \in G$ an operator $T_g$ on Fock space by
        \[
        T_g = \exp \left( \sum_{A,B,i} \bar c_A^i 
        (\ln g)_{AB}^{\vphantom{i}} c_B^i \right) .
        \]
In view of the multivaluedness of the logarithm, we need to demonstrate
that this operator is well-defined.  As the diagonalizable
supermatrices are dense in $G$, it suffices to do so for an element $g
\in G$ of the form $g = S \lambda S^{-1}$, where $\lambda$ is a
diagonal matrix containing the eigenvalues of $g$.  The eigenvalues of
$\ln g = S (\ln\lambda) S^{-1}$ are defined up to permutations and
shifts by integer multiples of $2\pi i$.  We follow the convention
that the Grassmann variables in the supermatrix $\ln g$ anticommute
with the fermionic operators of the set $\{ \bar c_A^i , c_B^j \}$.
It is then obvious that the transformation $c_A^i \mapsto
\gamma_A^i = \sum_B (S^{-1})_{AB}^{\vphantom{i}} c_B^i$, $\bar c_A^i
\mapsto {\bar\gamma}_A^i = \sum_B {\bar c}_B^i S_{BA}^{\vphantom{i}}$
is canonical.  Note $\sum {\bar c}_A^i (\ln g)_{AB}^{\vphantom{i}}
c_B^i = \sum \bar\gamma_A^i \gamma_A^i \ln\lambda_A$.  Now the number
operators $\hat n_A^i = {\bar\gamma}_A^i \gamma_A^i$ have integer
eigenvalues, so the ambiguity $\ln\lambda_A \to \ln\lambda_A + 2\pi
in$ is unobservable in $T_g = \exp \left( \sum_{A,i} \bar\gamma_A^i
\gamma_A^i \ln\lambda_A^{ \vphantom{i}} \right)$, and $T_g$ is indeed
well-defined.

We next claim that the mapping $g \mapsto T_g$ is a homomorphism of Lie
supergroups:
        \[
        T_{g} T_{h} = T_{gh} .
        \]
The proof goes as follows.  First, we use $T_{\exp X}$ to transform 
${\bar c}_A^i$ by
        \[
        {\bar c}_A^i \mapsto
        T_{\exp X}^{\vphantom{-1}} \ \bar c_A^i \ T_{\exp X}^{-1} = 
        \exp {\rm ad} (\bar c_B^j X_{BC}^{\vphantom{j}} c_C^j) \ 
        \bar c_A^i = \bar c_B^i (\exp X)_{BA}^{\vphantom{i}} ,
        \]
where ${\rm ad}(\hat A)\hat B := \hat A \hat B - \hat B \hat A$ stands 
for the commutator, and the summation convention is now back in force.
On setting $\exp X = g$ we get $T_g^{\vphantom{-1}} \bar c_A^i T_g^{-1} =
\bar c_B^i g_{BA}^{\vphantom{i}}$ and
        \[
        T_g^{\vphantom{-1}} T_h^{\vphantom{-1}} \bar 
        c_A^i T_h^{-1} T_g^{-1} = T_g^{\vphantom{-1}} \left( \bar c_B^i
        h_{BA}^{\vphantom{i}} \right) T_g^{-1} = 
        \bar c_C^i g_{CB}^{\vphantom{i}} h_{BA}^{\vphantom{i}}
        = \bar c_B^i (gh)_{BA}^{\vphantom{i}} = T_{gh}^{\vphantom{-1}} 
        \bar c_A^i T_{gh}^{-1} ,
        \]
from which we infer that the product 
$T_{gh}^{-1} T_g^{\vphantom{-1}} T_h^{\vphantom{-1}}$ 
commutes with ${\bar c}_A^i$.  For identical reasons, 
$T_{gh}^{-1} T_g^{\vphantom{-1}} T_h^{\vphantom{-1}}$ 
commutes also with $c_B^j$. 
Now, because the set of Fock operators $\{ {\bar c}_A^i , c_B^j \}$ act
irreducibly on Fock space, the vanishing commutator implies that
$T_{gh}^{-1} T_g^{\vphantom{-1}} T_h^{\vphantom{-1}}$  is a multiple
of the unit operator:
       \[
       T_{gh}^{-1} T_g^{\vphantom{-1}} T_h^{\vphantom{-1}}
       = f(g,h) \times {\bf 1} .
       \]
The second step is to show $f(g,h) \equiv 1$.  For that we use a 
formula\cite{helgason} for the differential of the exponential map
of any Lie algebra onto its Lie group:
        \begin{eqnarray}
        &&{d \over dt}\Big|_{t=0}
        e^{X + t{\dot X}} = e^X {\cal T}_X({\dot X}) , 
        \quad {\rm where} \nonumber \cr
        &&{\cal T}_X = \sum_{n=0}^\infty (-1)^n {{\rm ad}^n(X) \over
        (n+1)! } = {1 - e^{-{\rm ad}(X)} \over {\rm ad}(X)} \ .
        \nonumber
        \end{eqnarray}
This formula trivially extends to the case at hand, namely 
${\rm gl}(n|n)$, a Lie algebra with Grassmann structure\cite{berezin}.
The set where ${\cal T}_X$ has an inverse is dense in ${\rm gl}(n|n)$. 
We put $g = \exp X$, $h(t) = \exp tY$, and apply the formula for
the differential twice, the first time in the reverse order:
        \begin{eqnarray}
        {d \over dt}\Big|_{t=0}
        T_{\exp(X) \exp(tY)} &=& 
        {d \over dt}\Big|_{t=0}
        T_{\exp\left( X + {\cal T}_X^{-1} (tY) \right)} \nonumber \cr
        &=& {d \over dt}\Big|_{t=0}
        \exp {\bar c}_A^i \left( X + {\cal T}_X^{-1} (tY) \right)_{AB}
        c_B^i \nonumber \cr
        &=& \exp \left( {\bar c}_A^i X_{AB}^{\vphantom{i}} c_B^i \right)
        {d \over dt}\Big|_{t=0} \exp {\bar c}_A^i \left( 
        {\cal T}_X^{\vphantom{-1}} \circ {\cal T}_X^{-1}
        (tY) \right)_{AB} c_B^i , \nonumber \cr
        {\rm so} \qquad {d \over dt}\Big|_{t=0}
        T_{g h(t)} &=& {d \over dt}\Big|_{t=0} T_g T_{h(t)} .
        \nonumber
        \end{eqnarray}
By integrating the last equation one arrives at the desired
representation property, $T_{gh} = T_g T_h$. (q.e.d.)

The subgroup $H = {\rm Gl}(n_+ | n_+) \times {\rm Gl}(n_- | n_-)
\subset G$ consisting of elements $h = {\rm diag}(A,D)$ acts by
        \[
        T_{{\rm diag}(A,D)}
        = \exp \left( \bar c_{+a}^i (\ln A)_{ab}^{\vphantom{i}} c_{+b}^i
        + \bar c_{-a}^i (\ln D)_{ab}^{\vphantom{i}} c_{-b}^i \right) .
        \]
This subgroup stabilizes the vacuum:
        \[
        T_{{\rm diag}(A,D)} | 0 \rangle
        = | 0 \rangle \exp N \sum_a (-1)^{|a|+1} (\ln D)_{aa}
        = | 0 \rangle \ {\rm SDet} D^{-N} .
        \]
With $\mu( {\rm diag}(A,D) ) := {\rm SDet}D^{-N}$, we say that the 
vacuum carries a one-dimensional representation $\mu$ of $H$:
        \[
        T_h | 0 \rangle = | 0 \rangle \mu(h) , \quad {\rm and} \quad 
        \langle 0 | T_h^{-1} = \mu(h)^{-1} \langle 0 |  \qquad (h\in H).
        \]

After all these preliminaries we consider the generalized coherent
states $T_g | 0 \rangle$ $(g \in G)$ and, in particular, the operator 
$P$ defined by the integral
        \[
        P = \int_{M_{\rm B}\times M_{\rm F}} Dg_H \ 
        T_g^{\vphantom{-1}} | 0 \rangle \langle 0 | T_g^{-1} .
        \]
(Recall that $Dg_H$ denotes the $G$-invariant Berezin measure of the
coset space $G / H$, and the integration domain $M_{\rm B}\times
M_{\rm F}$ was specified in Sec.~\ref{sec:geometry}.)  The integrand
is a function on the coset space:
        \[
        T_{gh}^{\vphantom{-1}} | 0 \rangle \langle 0 | T_{gh}^{-1} = 
        T_g^{\vphantom{-1}} T_h^{\vphantom{-1}} 
        | 0 \rangle \langle 0 | T_h^{-1} T_g^{-1} = 
        T_g^{\vphantom{-1}} |0 \rangle \langle 0 | T_g^{-1} \quad (h \in H) ,
        \]
so the integral is well-defined.  The following calculation shows
that the operator $P$ commutes with $T_{g_0}$ for any $g_0 \in G$:
        \begin{eqnarray}
        T_{g_0} P &=& 
        \int Dg_H \ T_{g_0} T_g | 0 \rangle \langle 0 | T_g^{-1}
        \nonumber \\
        &=& \int Dg_H \ T_{{g_0}g} | 0 \rangle \langle 0 | T_g^{-1}
        \nonumber \\
        &=& \int Dg_H \ T_g | 0 \rangle \langle 0 | T_{g_0^{-1}g}^{-1}
        \nonumber \\
        &=& \int Dg_H \ T_g | 0 \rangle \langle 0 | T_g^{-1} T_{g_0}^{
        \vphantom{-1}} = P T_{g_0} ,
        \nonumber
        \end{eqnarray}
where the invariance of $Dg_H$ under translations $g \mapsto g_0 g$
was used.  (This invariance is crucial and is what determines $D(Z,
\tilde Z) = Dg_H$ in (\ref{HS}), up to multiplication 
by a constant.)  As was explained earlier, the action of the Lie
superalgebra of $G$ on the flavor sector is irreducible.  By Schur's
lemma then, the fact that $P$ commutes with all operators $T_{g_0}$
leads to the conclusion that $P$ is proportional to the identity on
the flavor sector.

Let us determine the constant of proportionality.  To that end, let
$\pi : G\to G / H$ be the canonical projection assigning group
elements $g$ to cosets $\pi(g) \equiv g H$.  We make a decomposition
$g = s(\pi(g)) h(g)$, where $s(\pi(g))$ takes values in $G$ and $h(g)$
takes values in $H$.  (Such a decomposition fixes a ``choice of
gauge''\cite{mstone,mrz_iqhe}, and $s : G / H \to G$ is called a
(local) section of the bundle $\pi : G \to G / H$.)  In this way we
get
        \begin{eqnarray}
        T_g | 0 \rangle &=& T_{s(\pi(g))} T_{h(g)} | 0 \rangle
        = T_{s(\pi(g))} | 0 \rangle \ \mu(h(g)) ,
        \nonumber \\
        \langle 0 | T_g^{-1} &=& \mu(h(g))^{-1} \ 
        \langle 0 | T_{s(\pi(g))}^{-1} , \quad
        {\rm and} \quad T_g^{\vphantom{-1}} | 0 \rangle \langle 0 | T_g^{-1} = 
        T_{s(\pi(g))}^{\vphantom{-1}} 
        | 0 \rangle \langle 0 | T_{s(\pi(g))}^{-1} .
        \nonumber
        \end{eqnarray}
Following the main text we put $g = \left( \mymatrix{A &B\cr C &D\cr} 
\right)$ and parameterize cosets $\pi(g)$ by the pair $(Z = BD^{-1},
\tilde Z = CA^{-1})$.  For $s$ we choose
        \begin{eqnarray}
        s(\pi(g)) \equiv s(Z,\tilde Z) &=& 
        \pmatrix{ (1-Z\tilde Z)^{-1/2} &Z(1-\tilde Z Z)^{-1/2}\cr 
        \tilde Z(1-Z\tilde Z)^{-1/2} &(1-\tilde Z Z)^{-1/2} \cr}
        \nonumber \\
        &=& \pmatrix{1 &Z\cr 0 &1\cr}
        \pmatrix{ (1-Z\tilde Z)^{+1/2} &0\cr 0 &(1-\tilde Z Z)^{-1/2}}
        \pmatrix{1 &0\cr \tilde Z &1\cr} .
        \nonumber
        \end{eqnarray}
The first line shows that this is a valid choice of section, i.e. we 
indeed have $Z = BD^{-1}$ and $\tilde Z = C A^{-1}$.  Using the second 
line to translate $s(Z,\tilde Z)$ into an operator on Fock space, we obtain
        \begin{eqnarray}
        T_{s(Z,\tilde Z)} | 0 \rangle &=& 
        \exp \left( \bar c_{+a}^i Z_{ab}^{\vphantom{i}} c_{-b}^i \right)
        \exp \Big( {\textstyle{1\over 2}} \bar c_{+a}^i
        \ln (1 - Z\tilde Z)_{ab}^{\vphantom{i}} c_{+b}^i 
        \nonumber \\ 
        &&\hspace{4cm} - {\textstyle{1\over 2}} \bar c_{-a}^i 
        \ln (1-\tilde Z Z)_{ab}^{\vphantom{i}} c_{-b}^i \Big) \exp \left ( 
        \bar c_{-a}^i \tilde Z_{ab}^{\vphantom{i}} c_{+b}^i \right) | 0 \rangle
        \nonumber \\
        &=& \exp \left( \bar c_{+a}^i Z_{ab}^{\vphantom{i}} c_{-b}^i \right)
        | 0 \rangle \ {\rm SDet}(1-\tilde Z Z)^{N/2} =: | Z \rangle ,
        \quad {\rm and} \nonumber \\
        \langle 0 | T_{s(Z,\tilde Z)}^{-1} &=& 
        \langle 0 | T_{s(-Z,-\tilde Z)} = 
        {\rm SDet}(1-\tilde Z Z)^{N/2} \langle 0 |
        \exp \left( - \bar c_{-a}^i \tilde Z_{ab}^{\vphantom{i}} c_{+b}^i 
        \right) =: \langle Z | .
        \nonumber
        \end{eqnarray}
The expression for the invariant measure $Dg_H$ in the
coordinates $Z, \tilde Z$ is $D(Z,\tilde Z)$, so
        \[
        P = \int_{M_{\rm B}\times M_{\rm F}} Dg_H \ 
        T_g^{\vphantom{-1}} | 0 \rangle \langle 0 | T_g^{-1} 
        = \int D(Z,\tilde Z) \ T_{s(Z,\tilde Z)}^{\vphantom{-1}} 
        | 0 \rangle \langle 0 | T_{s(Z,\tilde Z)}^{-1} = 
        \int D(Z,\tilde Z) | Z \rangle \langle Z | .
        \]
Taking the vacuum expectation value of $P$ we find
        \[
        \langle 0 | P | 0 \rangle = \int D(Z,\tilde Z) \ \langle 0 | 
        Z \rangle \langle Z | 0 \rangle = \int D(Z,\tilde Z)
        \ {\rm SDet}(1-\tilde Z Z)^N = \int D\mu_N(Z,\tilde Z) = 1 ,
        \]
by our choice of normalization, see Sec.~\ref{sec:HS}. It follows that 
the constant of proportionality between $P$ and the unit operator on the 
flavor sector is unity.  Because $\langle Z |$ vanishes on states that 
are not color-singlets, $P$ annihilates all states outside the flavor 
sector.  Hence $P$ {\it projects Fock space onto the flavor sector}.

The final ingredient we need are the Bose-Fermi coherent states
        \[
        \exp \left( \bar c_{+a}^i \psi_{+a}^i + \bar\psi_{-b}^i c_{-b}^i 
        \right) | 0 \rangle .
        \]
Note that the ordering in the exponential matters as the Grassmann
variables are taken to anticommute with the fermionic operators: $\bar
c_{+a}^i \psi_{+a}^i = (-1)^{|a|} \psi_{+a}^i \bar c_{+a}^i$, as
before.  These Bose-Fermi coherent states span the entire Fock space
(or, rather, the ``Grassmann envelope''\cite{berezin} thereof). 
They can be projected on the flavor (or color-neutral) sector by making
a unitary rotation in color space, $c_A^i \mapsto c_A^j U^{ji}$, 
$\bar c_A^i \mapsto {\bar U}^{ji} \bar c_A^j$, and averaging over all such
rotations.  Therefore the operator $P$ that projects on the flavor sector
acts on Bose-Fermi coherent states as
        \[
        P \exp \left( \bar c_{+a}^i \psi_{+a}^i + 
        \bar\psi_{-b}^i c_{-b}^i \right) | 0 \rangle 
        = \int_{{\rm U}(N)} dU \ 
        \exp \left( {\bar U}^{ji} \bar c_{+a}^j \psi_{+a}^i +
        \bar\psi_{-b}^i c_{-b}^j U^{ji} \right) | 0 \rangle .
        \]
With all these tools in hand, the statement (\ref{HS}) is
proved by the following computation:
        \begin{eqnarray}
        &&\int D\mu_N(Z,\tilde Z) \ \exp \left(
        \bar\psi_{+a}^i Z_{ab}^{\vphantom{i}} \psi_{-b}^i +
        \bar\psi_{-b}^i \tilde Z_{ba}^{\vphantom{i}} \psi_{+a}^i \right)
        \nonumber \\
        &=& \int D\mu_N(Z,\tilde Z) \ \langle 0 |
        \exp \left( \bar\psi_{+a}^i c_{+a}^i - 
        \bar c_{-b}^i \psi_{-b}^i \right) 
        \exp \left( \bar c_{+a}^i Z_{ab}^{\vphantom{i}}
         c_{-b}^i \right) | 0 \rangle 
        \nonumber \\
        &&\hspace{2cm} \times \langle 0 | 
        \exp \left( - \bar c_{-b}^i \tilde Z_{ba}^{\vphantom{i}} 
        c_{+a}^i \right)
        \exp \left( \bar c_{+a}^i \psi_{+a}^i + \bar\psi_{-b}^i c_{-b}^i 
        \right) | 0 \rangle 
        \nonumber \\
        &=& \langle 0 |         
        \exp \left( \bar\psi_{+a}^i c_{+a}^i - 
        \bar c_{-b}^i \psi_{-b}^i \right) \left( \int D(Z,\tilde Z)
        | Z \rangle \langle Z | \right)
        \exp \left( \bar c_{+a}^i \psi_{+a}^i + 
        \bar\psi_{-b}^i c_{-b}^i \right) | 0 \rangle 
        \nonumber \\
        &=& \langle 0 |         
        \exp \left( \bar\psi_{+a}^i c_{+a}^i - 
        \bar c_{-b}^i \psi_{-b}^i \right) {P}
        \exp \left( \bar c_{+a}^i \psi_{+a}^i + 
        \bar\psi_{-b}^i c_{-b}^i \right) | 0 \rangle 
        \nonumber \\
        &=& \int_{{\rm U}(N)} dU \ 
        \langle 0 | \exp \left( \bar\psi_{+a}^i c_{+a}^i - 
        \bar c_{-b}^i \psi_{-b}^i \right) 
        \exp \left( \bar c_{+a}^i U^{ij} \psi_{+a}^j +
        \bar\psi_{-b}^i \bar U^{ji} c_{-b}^j \right) | 0 \rangle 
        \nonumber \\
        &=& \int_{{\rm U}(N)} dU \ \exp \left( 
        \bar\psi_{+a}^i U^{ij} \psi_{+a}^j +    
        \bar\psi_{-b}^i \bar U^{ji} \psi_{-b}^j \right) .
        \nonumber
        \end{eqnarray}
The first equality sign is an elementary consequence of the
commutation relations of the Fock operators $c, \bar c$ and their
action on the vacuum $| 0 \rangle$.  The second and third equality
signs recognize the projector $P$ on the flavor sector.  The fourth
equality sign implements $P$ by averaging over all unitary rotations
in color space.  The last equality sign is elementary again.

\end{document}